\documentclass[authoryear,12pt]{article}
\usepackage{amsmath}
\usepackage{graphicx} 
\usepackage{amsfonts}%
\usepackage{amssymb}
\usepackage{pdflscape}
\usepackage{floatrow}
\usepackage{times}
\usepackage{authblk}
\usepackage{blindtext}
\usepackage[]{apacite}
\RequirePackage[colorlinks,citecolor=blue,urlcolor=blue]{hyperref}
 \usepackage{natbib}
\usepackage{floatrow}
\allowdisplaybreaks
 \allowbreak

\usepackage{subcaption}
\usepackage{longtable}
\usepackage{rotating}
\allowdisplaybreaks

\usepackage{enumitem}
\usepackage{amsthm}
\usepackage{setspace}
\usepackage{algorithm}
\usepackage{algpseudocode}
\usepackage{caption}
\usepackage{blindtext}
\usepackage{multirow}
\newfloatcommand{capbtabbox}{table}[][\FBwidth]
\usepackage[margin=1in]{geometry}
\theoremstyle{plain}
\newtheorem{theorem}{Theorem}

\newcommand{\bbeta}{ \mbox{\boldmath $ \beta $} }

\newcommand{\bone}{\textbf{1}}

\newcommand{\bA}{\textbf{A}}

\newcommand{\bB}{\textbf{B}}

\newcommand{\bC}{\textbf{C}}

\newcommand{\bE}{\textbf{E}}

\newcommand{\bF}{\textbf{F}}

\newcommand{\bs}{\textbf{s}}

\newcommand{\bw}{\textbf{w}}

\newcommand{\bx}{\textbf{x}}

\newcommand{\by}{\textbf{y}}
\newcommand{\bY}{\textbf{Y}}

\newcommand{\R}{\mathbb{R}}
\renewcommand{\S}{\mathbb{S}}

\title{Modeling Daily Seasonality of Mexico City Ozone using Nonseparable Covariance Models on Circles Cross Time}

\author[1]{Philip A. White \thanks{Corresponding Author: paw27@duke.edu}}
\affil[1]{Department of Statistical Science, Duke University, Durham, NC, USA}
\author[2]{Emilio Porcu \thanks{emilio.porcu@newcastle.ac.uk}}
\affil[2]{School of Mathematics, Statistics and Physics; Chair of Spatial Analytics Methods (SAM),  Newcastle University, UK}

\begin{document}

\maketitle

\begin{abstract}
Mexico City tracks ground-level ozone levels to assess compliance with national ambient air quality standards and to prevent environmental health emergencies. Ozone levels show distinct daily patterns, within the city, and over the course of the year. To model these data, we use covariance models over space, circular time, and linear time. We review existing models and develop new classes of nonseparable covariance models of this type, models appropriate for quasi-periodic data collected at many locations. With these covariance models, we use nearest-neighbor Gaussian processes to predict hourly ozone levels at unobserved locations in April and May, the peak ozone season, to infer compliance to Mexican air quality standards and to estimate respiratory health risk associated with ozone. Predicted compliance with air quality standards and estimated respiratory health risk vary greatly over space and time. In some regions, we predict exceedance of national standards for more than a third of the hours in April and May. On many days, we predict that nearly all of Mexico City exceeds nationally legislated ozone thresholds at least once. In peak regions, we estimate respiratory risk for ozone to be 55\% higher on average than the annual average risk and as much at 170\% higher on some days.

\end{abstract}

\section{Introduction}\label{sec:intro}

Ground-level ozone is linked with short- and long-term health risks in general \citep[see, e.g.,][]{lippmann1989,salam2005,bell2006,weschler2006} and in Mexico City specifically \citep[see][]{mage1996, romieu1996, hernandez1997, loomis1999, bravo2002, barraza2008, riojas2014}. In statistics, many have studied daily ozone levels \citep[see, e.g.,][]{sahu2007, berrocal2010, huang2018}. However, because short-term spikes in ozone levels are associated with increased respiratory health risk, some argue that using finer time scales, normally hourly, and treating time as continuous, is essential to quantify short-term health risks \citep{chiogna2011,arisido2016}.

The Mexico City ozone monitoring data presented here consist of hourly measurements from 24 stations from April and May, Mexico City's peak ozone season. Of primary interest in this analysis are predicted ozone levels at unmonitored locations and associated compliance to national ambient air quality standards and respiratory health risk. For compliance, we discuss the probability of exceeding nationally legislated limits, 95 parts per billion (ppb) for \emph{hourly} ozone and 70 ppb for \emph{eight-hour average} ozone \citep{nom14b}. For health outcomes, we use methods in \cite{chiogna2011} and compare respiratory health risks during the peak ozone season to the average ozone risk over 2017. Our model must enable these goals. First, our model must allow space-time predictions, accounting for daily seasonality or periodicity. With these predictions, we make probabilistic assessments of compliance and respiratory health risk. 

\begin{figure}[H]
\begin{center}
\includegraphics[width=0.4\textwidth]{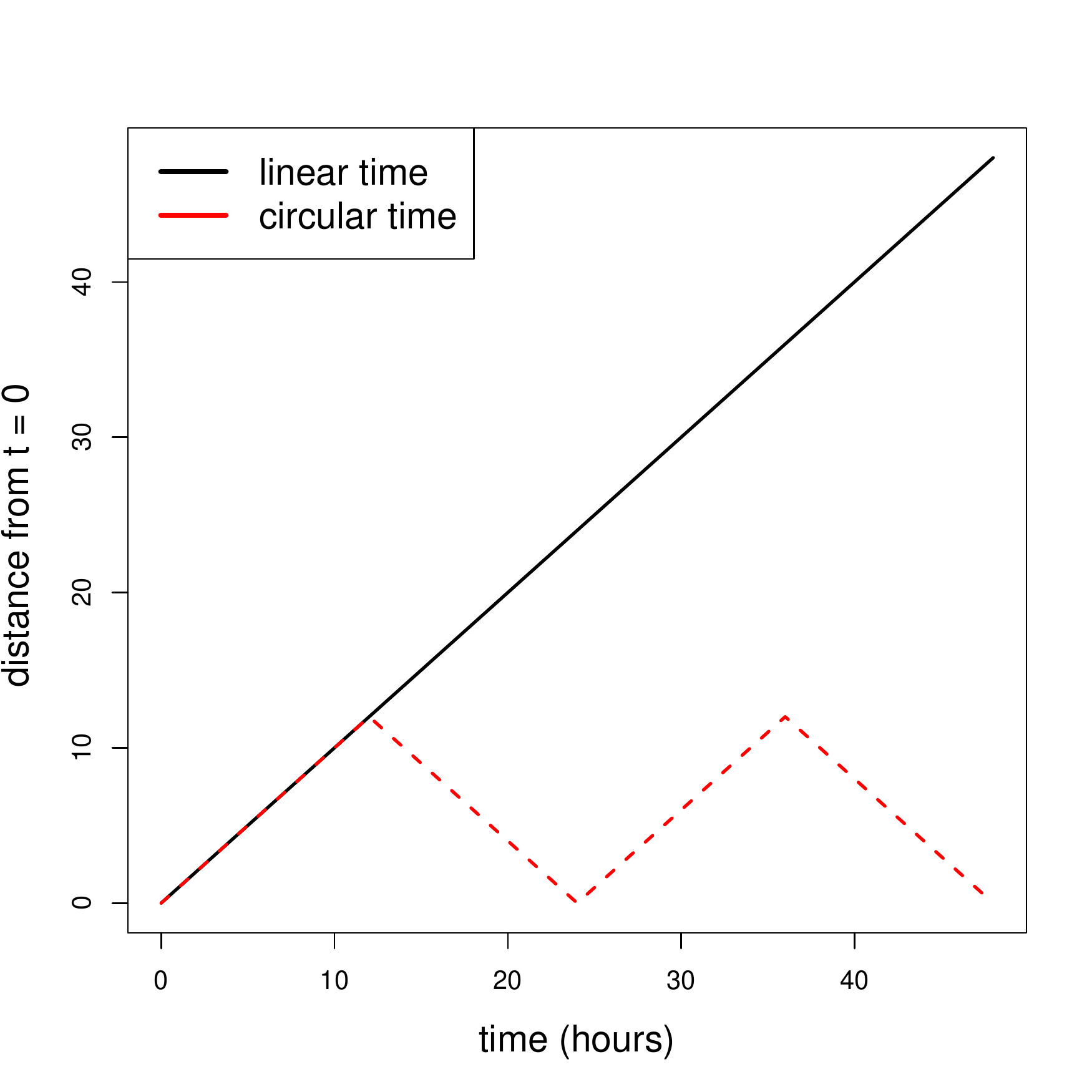}
\end{center}

\caption{This represents our two ideas of temporal differences, using both raw (linear) differences and circular differences (angular difference) in time. }\label{fig:circle_linear}
\end{figure}

Here, we provide three primary contributions. To account for the daily pattern of ozone in the Mexico City data, we envision time as a quantity that lies on a circle and the real line (i.e. a 24-hour clock and linear time) (see Figure \ref{fig:circle_linear} for illustration). To develop appropriate models for this, we extend the current literature on covariance functions for circles cross time and demonstrate that examples from these new classes have better predictive performance for the Mexico City data than current alternatives. Then, we discuss modeling details for Vecchia models that rely on neighbor selections for model fitting and prediction \citep{vecchia1988}. This problem has not been addressed in the literature for periodic data and is not as simple as covariance models that decay monotonically over space and time. Lastly, we use this modeling framework to assess compliance with air quality standards and respiratory health risks.

Periodic or seasonal data, as we define it, rise and fall with some fixed period or alternatively has covariance that follows a periodic pattern. With this loose definition of seasonality, we include covariance functions that dampen as a function of time while exhibiting seasonality, for which we borrow the term \emph{quasi-periodic} from \cite{solin2014}. The Mexico City ozone data exhibits this type of pattern, as we discuss in Section \ref{sec:data}.

Treating seasonal data in discrete time is common \citep[for reference, see][]{west1997,prado2010,shumway2017}. These types of models have even been applied to Mexico City ozone levels \citep{huerta2004,white2018a} but were used with different goals in mind. Because prediction at unobserved locations and potentially unobserved times is essential for our modeling goals, we argue for continuous space and continuous time models.

Although the discussion on positive-definite functions on spheres dates to \cite{schoenberg1942}, we take the work of \cite{gneiting2013} as our starting point. Extensions of this work to spheres cross time include \cite{porcu2016} and \cite{white2018b}, while \cite{shirota2017} developed similar covariance models for circles cross time to account for time-of-day effects for modeling crime event data. We further this discussion by proposing new classes of nonseparable covariance functions specific for circles cross time and by extending current covariance classes to this setting. For discussion on tests for correlation functions on circles, see \cite{gneiting1998}.

For scalable Gaussian models, we use methods that induce sparsity in the inverse covariance matrix. First proposed by \cite{vecchia1988} and then extended by \cite{stein2004}, \cite{stein2005}, \cite{bev}, \cite{gramacy2015}, \cite{datta2016a}, and \cite{katzfuss2017}, these methods reduce computational expense by assuming conditional independence given a neighborhood (conditioning) set. We refer to these approaches as Vecchia models and focus specifically on the nearest-neighbor Gaussian process \citep{datta2016a}. 

Discussion about periodic data for neighbor-based approaches is lacking in the literature. \cite{stein2005} considers daily data where periodicity is not present and conditions on most recent times. \citet{datta2016c} work on the same time scale and argue using dynamic neighborhood sets that select the most correlated neighbors; however, their approaches do not easily generalize to our setting. In this article, we open this discussion for quasi-periodic covariance models.

We continue by discussing the Mexico City pollution monitoring data in Section \ref{sec:data}. Then, we discuss appropriate covariance classes for these data and present new covariance classes for circles cross time in Section \ref{sec:covar}. Using these and other covariance classes, we discuss modeling details using Bayesian nearest-neighbor Gaussian processes models in Section \ref{sec:modeling}. We motivate this discussion in the Online Supplement using a single time-series. We present the results of our analysis in Section \ref{sec:space_time}, addressing compliance to Mexican ambient air quality standards and respiratory health risks associated with ground-level ozone. Lastly, we give concluding remarks and discuss future extensions in Section \ref{sec:conc}.

\section{Mexico City Ozone Monitoring Data}\label{sec:data}

In this dataset, we have hourly ozone measurements for April and May of 2017 at $N_s = 24$ monitoring stations across Mexico City, Mexico. At each station, we have measurements at $N_t = 1464$ times, giving $N = 35136$ observations in total\footnote{Although ozone levels are given hourly, these hourly quantities are derived is an average of the 60 minute-by-minute measurements. Missing measurements were imputed prior to receiving the data using the measurements at the nearest station within the same region. If no simultaneous measurements in the same region were available, then the missing measurement was filled in using the nearest station in a different region.}. Relative humidity (RH) and temperature (TMP) are also measured hourly at the same 24 stations, and these are used as explanatory variables.

Ozone is formed by volatile nitrous oxides exposed to heat and sunlight \citep{sillman1999}; thus, ozone levels peak during the warmest hours of the day. Ozone levels are held in check by rain that clears out pollutants. To use heat and rain as explanatory variables, we include hourly temperature and relative humidity as covariates for models in Sections \ref{sec:space_time} and the Online Supplement. 

\begin{figure}[H]
  \begin{center}
      \includegraphics[width=.75\textwidth]{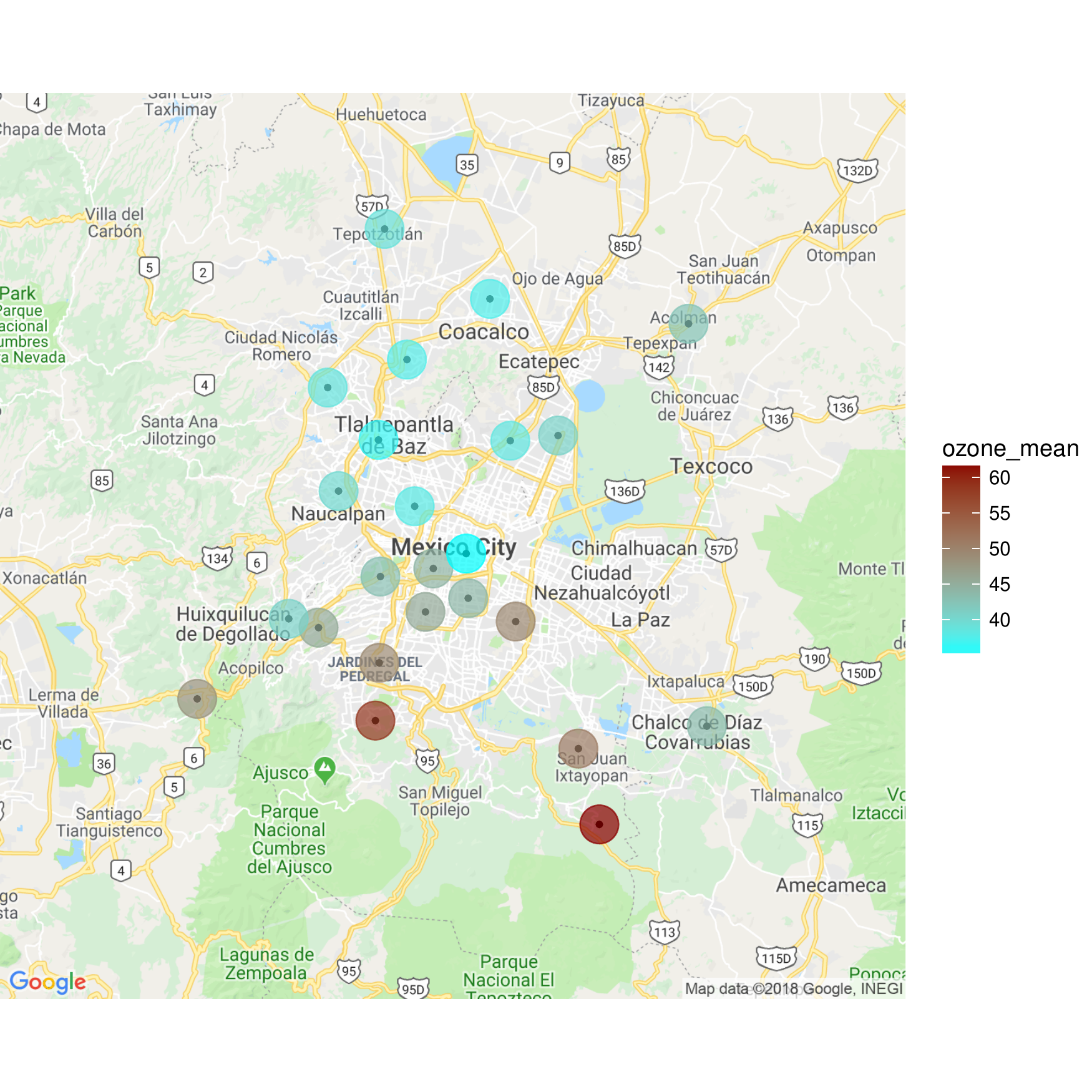}
  \end{center}
  \vspace{-12mm}
       \caption{Station locations with mean ozone coded by the color of the point with low to high ozone indicated by cyan to red.}\label{fig:station_loc}
\end{figure}

First, we look at variability in ozone levels across the city. To do this, we plot station locations with their mean ozone levels in Figure \ref{fig:station_loc} using the ggmap package in R \citep{kahle2013}. In general, ozone decreases as we move northward; however, this trend is not uniform, as central Mexico City has the lowest ozone values. Peak ozone levels in the south are largely explained by a wind corridor that flows northeast to southwest, moving ozone and ozone precursors produced along this wind path to southern parts of Mexico City. Additionally, ozone is often trapped in the south by mountains on Mexico City's southwest boundary.

\begin{figure}[H]
\begin{center}
   \begin{subfigure}[b]{.32\textwidth}
\includegraphics[width=\textwidth]{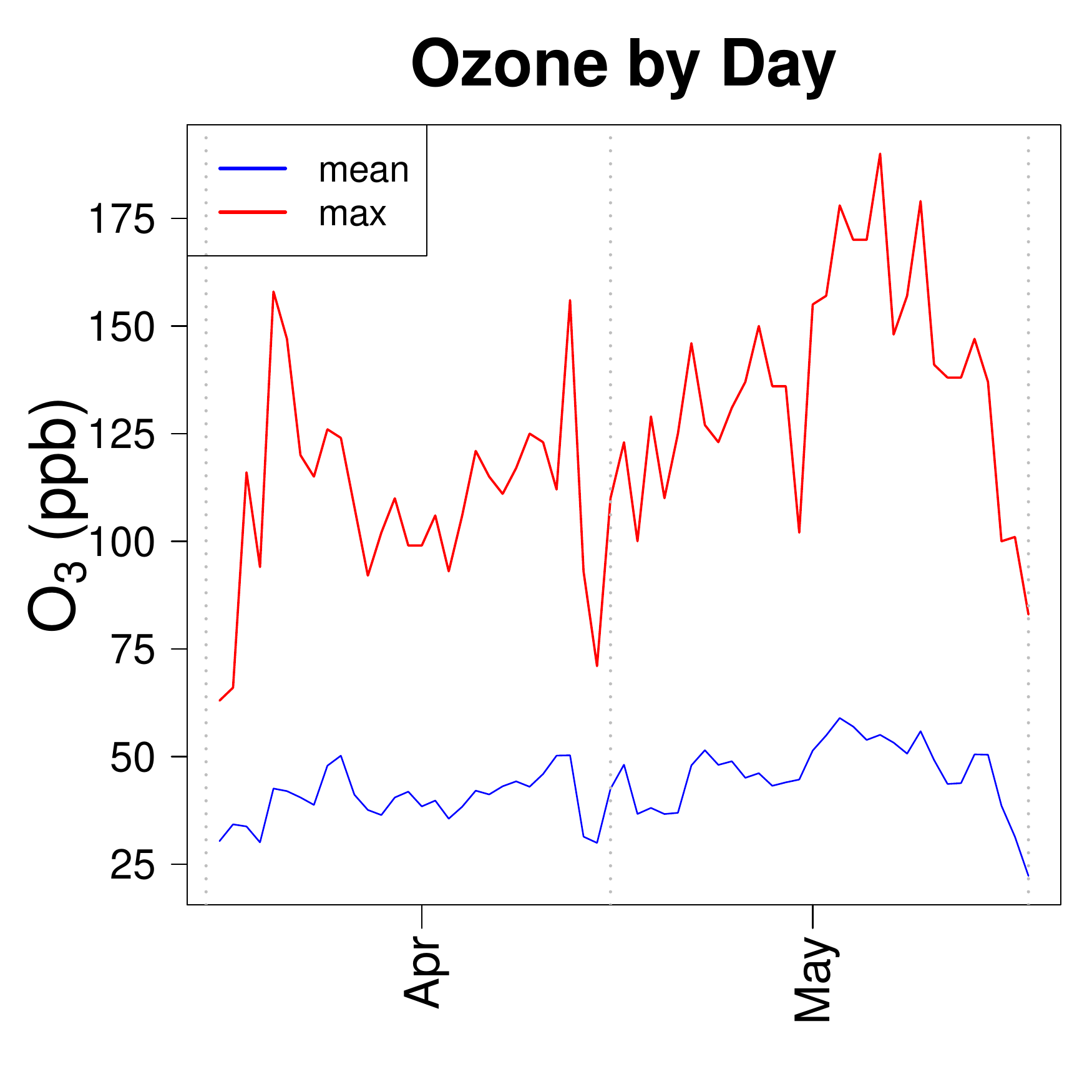}
      \subcaption{Mean and max over days}\label{fig:mex_mean_max_day}
   \end{subfigure}
   \begin{subfigure}[b]{.32\textwidth}
\includegraphics[width=\textwidth]{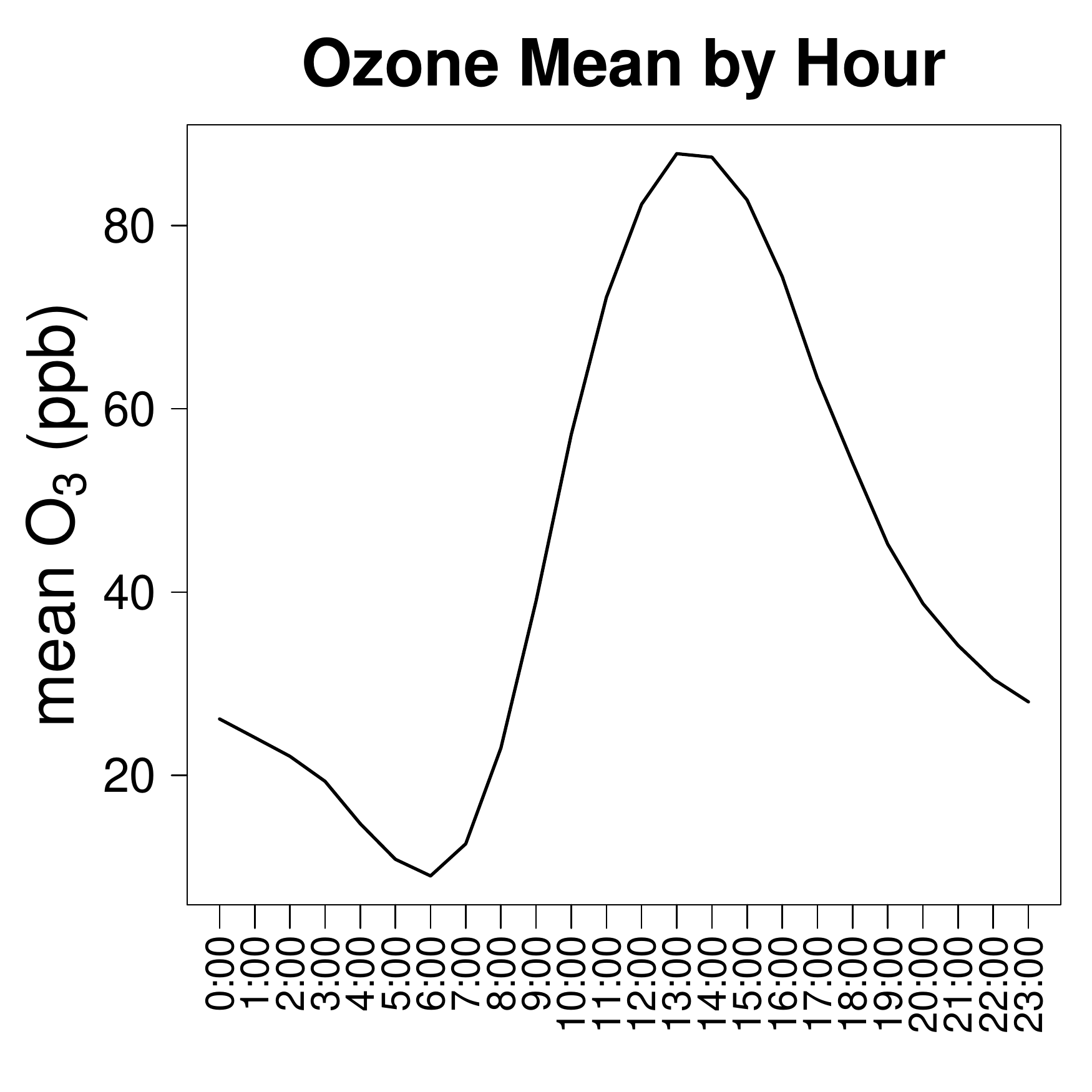}
      \subcaption{Mean ozone over hours}\label{fig:mex_hour}
   \end{subfigure}
   \begin{subfigure}[b]{.32\textwidth}
\includegraphics[width=\textwidth]{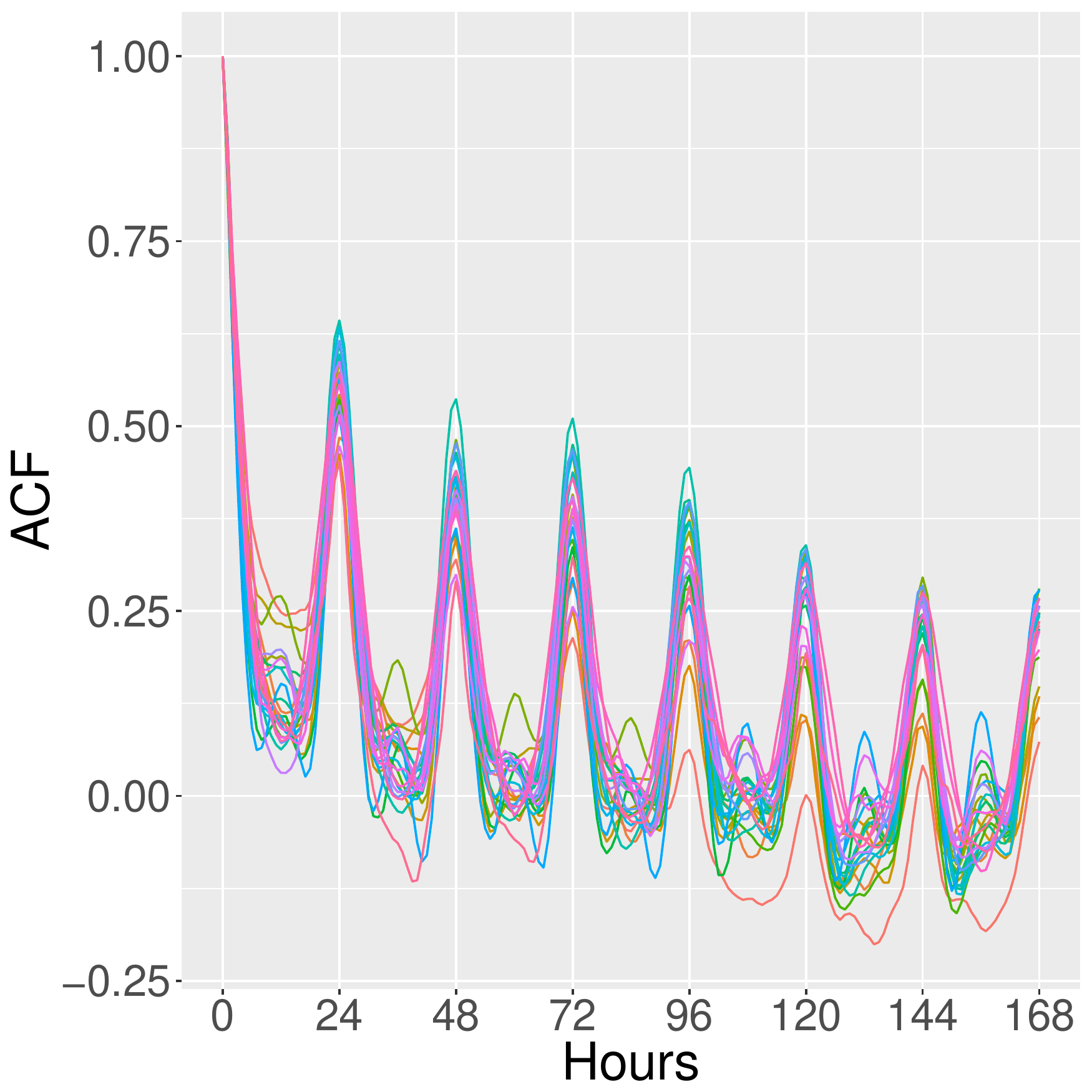}
      \subcaption{ACF for ozone }\label{fig:acf_o3_mex}
   \end{subfigure}
   \end{center}

\caption{(Left) Mean and maximum ozone levels for all stations each day (Center) Mean ozone over the hour of the day, averaging over days and stations (Right) Autocorrelation function (ACF) of the model residuals for each station using relative humidity and temperature as explanatory variables. Each color represents the ACF of each station.}
\label{fig:mex_means}
\end{figure}

We explore how the data vary over time, Figure \ref{fig:mex_mean_max_day} displays the mean and maximum ozone concentration for each day, showing that ozone concentrations peak in May. Figure \ref{fig:mex_hour} plots ozone averages as a function of hour of the day and demonstrates a clear peak in ozone levels around 2:00 or 3:00 p.m.
 
To examine whether the daily pattern in ozone can be explained by temperature and humidity, we fit a linear model to ozone using relative humidity and temperature as covariates. Then, we examine the autocorrelation of the model residuals for each of the 24 locations (See Figure \ref{fig:acf_o3_mex}). For each site, the temporal autocorrelation pattern peaks every 24 hours but decays overall as a function of time. Thus, a purely periodic function or purely decaying covariance model would be insufficient for these data (we demonstrate this empirically in both our temporal analysis in the Online Supplement and Section \ref{sec:space_time}). These plots motivate our theoretical discussion in Section \ref{sec:covar}.

\section{Covariance Models }\label{sec:covar}

\subsection{Covariance Modeling Approach}

\def\x{\boldsymbol{x}}
\def\h{\boldsymbol{h}}

We start with the notation and background needed for a neat exposition of our results. We denote $\S^1 \subset \R^2$ as the unit circle and use the mapping $\theta: \S^1 \times \S^1 \to [0,\pi]$ as the angle (great circle distance) between any two points on $\S^1$. We use the space $\S^1$ to capture the circular nature of time exhibited in daily patterns in ozone levels. We use ${\cal D}$ as the spatial domain, a projection of a small portion of the sphere onto a plane. Throughout, we denote $\{Y( \x,t), \; \x \in {\cal D}, \; t \in \R \}$ as a weakly stationary space-time Gaussian random field. 

The assumption of Gaussianity implies that finite dimensional subsets of $Y$ are uniquely determined by its mean and the covariance function $K: \R^2 \times \R \to \R$, defined as  
$ K(\h,u)=  {\rm cov} \left ( Z(\x,t ), Z(\x',t') \right ),$
where $\h =\x -\x'$ and $u=t-t'$ are the spatial and the temporal lag, respectively.  
Our novel contribution is to use the temporal lag to account for differences in raw time ($u$ itself) and time-of-day, which we represent as the angle $\theta$ on a circle.
Thus, we assume 
\begin{equation}
\label{assumption} K(\h,u) = C(\| \h \|, \theta, |u|), \qquad \h \in \R^2, \theta \in [0,\pi], u \in \R,
\end{equation}
for some mapping $C: [0,\infty) \times [0,\pi] \times [0,\infty) \to \R$.
Constructing covariance functions of the type (\ref{assumption}) is challenging, and we propose covariance functions that are {\em partially} nonseparable. As an abuse of notation, we write $h$ for $\|\h\|$ and $u$ for $|u|$ and consider the following nonseparabilities:  
\begin{description}
\item[(A) $({\cal D},\S^1)-$ nonseparability.] Here, $C(h,\theta,u)= C_{1}(h,\theta) C_{2}(u)$; 
\item[(B) $({\cal D},\R)-$ nonseparability,] obtained when $C(h,\theta,u)= C_{3}(h,u) C_{4}(\theta)$;
\item[(C) $(\S^1,\R)-$ nonseparability,] where $C(h,\theta,u)= C_{5}(\theta,u) C_{6}(h)$.
\end{description}
Apparently, $C_i$, $i=1,\ldots,6$ are covariance functions on their respective subspaces. For the Mexico City data, we also consider complete separability, $C(h,\theta,u)= C_6(h) C_4(\theta) C_2(u)$, to motivate the need of nonseparable models. 
We detail some approaches for constructions (\bA), (\bB) and (\bC). 

For ease of illustration, we first define Gneiting functions, ${\cal G}_d: [0,\infty)^2 \to \R_+$, according to \cite{gneiting2002}, as follows:
\begin{equation}
\label{gneiting_form} {\cal G}_d(x,t):= \frac{1}{\psi(t)^{d/2}} \varphi \left ( \frac{x}{\psi(t^2)} \right ), \qquad x,t \ge 0. 
\end{equation}
Following \cite{porcu2016}, we define the modified Gneiting class, ${\cal P}: [0,\infty) \times [0,\pi] \to \R$, as 
\begin{equation}
\label{gneiting_form_2} {\cal P}(x,t):= \frac{1}{\psi_{[0,\pi]}(t)^{1/2}} \varphi \left ( \frac{x}{\psi_{[0,\pi]}(t)} \right ), \qquad x\ge 0 ,t \in [0,\pi].
\end{equation}
For both classes, the function $\varphi: [0,\infty) \to \R_+$ is completely monotonic; that is, $\varphi$ is infinitely differentiable on $(0,\infty)$, satisfying $(-1)^n \varphi^{(n)}(t) \ge 0$, $n \in \mathbb{N}$. The function $\psi$ is strictly positive and has a completely monotonic derivative. Here, $\psi_{[0,\pi]} $ denotes the restriction of $\psi$ to the interval $[0,\pi]$. We provide selections for $\varphi(\cdot)$ and $\psi(\cdot)$ in Tables 4 and 5 of the Online Supplement. The class ${\cal G}_d$ cannot not capture seasonality or periodicity but is appropriate for strategy (\bB), i.e. $C_3(h,u)={\cal G}_2(h,u)$ (see Section \ref{app:space_time_ex} for examples). We limit our selections of this type to ${\cal G}_d$, although many other examples exist in the literature. For strategy (\bA), arguments in Theorem 2 in \cite{white2018b} ensure $C_1(h,\theta)= {\cal  G}_2(\theta,h^2)$ to provide a valid covariance function, and Theorem 1 in \cite{porcu2016} shows that $C_1(h,\theta)= {\cal P}(h^2,\theta)$ is a valid covariance function. These classes can also be used for $C_5(\theta,u)$ in approach $(\bC)$. Examples for $C_1(h,\theta)$ and $C_5(\theta,u)$ are given in Appendix \ref{app:time_ex}.

The marginal covariance $C_2$ is taken from the Mat{\'e}rn class ${\cal M}_{\alpha,\nu}: [0,\infty) \to \R_+$, defined as \begin{equation}
\label{matern} {\cal M}_{\alpha,\nu}(t) = \frac{2^{1-\nu}}{\Gamma(\nu)} \left ( \frac{t}{\alpha}\right )^{\nu} {\cal K}_{\nu } \left ( \frac{t}{\alpha}\right ), \qquad \alpha, \nu >0,
\end{equation}
where ${\cal K}_{\nu}$ is the MacDonald function \citep{grad}. Well known arguments show that $C_2(u)= {\cal M}_{\alpha,\nu}(u)$ is a valid choice \citep{stein1999} for any $\alpha,\nu > 0$. The exponential model on $\R$, ${\cal M}_{\alpha,1/2}(t)= \exp(-t/\alpha)$, induces conditional independence between random effects given the time immediately previous, regardless of spacing (see the Online Supplement for some discussion). A severe restriction on the parameter $\nu$ is required for $C_4(\theta) = {\cal M}_{\alpha,\nu}(\theta)$ to be a valid covariance \citep[see][]{gneiting2013}: $\nu \in (0,1/2]$. The Mat{\'e}rn function can also be used to generate valid covariance $C_6(h)= {\cal M}_{\alpha,\nu}(h)$ on any $d$-dimensional Euclidean space \citep{stein1999}. Choices for the functions $C_5(\theta,u)$ for construction {\bf (C)} are the core of our theoretical contributions. 

\subsection{Covariance Functions for Circles Cross Time}\label{sec:covar_theory1}

Literature on periodic covariance functions is sparse in general, with some exceptions. For fixed period $P$, \cite{mackay1998} \emph{embeds} the temporal mapping, $ t \mapsto \Big  (\sin \left (\frac{P}{2\pi}t  \right ),\cos  \left (\frac{P}{2\pi}t \right ) \Big )$, in the squared exponential correlation function to obtain a periodic covariance model. 
\cite{rasmussen2006} propose quasi-periodic covariance functions by taking the product of periodic and decaying covariance functions. 
\cite{solin2014} show an explicit link between some periodic and separable quasi-periodic Gaussian process specifications and state-space models.
Most discussion about periodic covariance models is, however, limited to separable models like those discussed. To provide more general nonseparable covariance models for periodic data, we turn to models on circles cross time.

For hourly data, we map hour-of-day onto a 24-hour clock or circle. As discussed, there are rich covariance classes on spheres and spheres cross time \citep[e.g.,][]{gneiting2013,porcu2016,white2018b}. \cite{shirota2017} discuss nonseparable covariance classes on $\S^1  \times \R^2 $ (circles cross space) for crime event data using log-Gaussian Cox processes. However, for our data, purely circular classes from \cite{gneiting2013} and \cite{shirota2017} are too limited because they do not dampen over time.

The result we present is motivated by autocorrelation patterns in Figure \ref{fig:acf_o3_mex} in Section \ref{sec:data}, exhibiting both periodicity and decay. Specifically, we propose two classes on nonseparable covariance functions on $\mathbb{S}^1 \times \mathbb{R}$. For these classes, we define the variogram of an intrinsically stationary process, $\gamma: \R \to [0,\infty)$, as $\gamma(u) = \frac{1}{2}{\rm var} \left ( Z(t+u)-Z(t) \right ) $, $t,u \in \R$. We also recall that, for any covariance function on $\S^1 \times \R$, the ratio $\rho(\theta,u)= C(\theta,u)/C(0,0)$ is a correlation function. Lastly, we define $\sinh$ as the hyperbolic sine function. 
\begin{theorem}\label{thm:circle1} Let $\gamma: \R \to \R_+$ be a variogram and $\rho: \R \to \R$ be a correlation function. 
\begin{description}
\item[(I)] Let $C$ be defined as 
\begin{equation} \label{emi1}
C(\theta,u) = \frac{1}{\gamma(u)} + \frac{\pi}{2} \frac{\sinh \left[ \sqrt{\gamma(u) } (\pi - \theta) \right] }{\sinh \left[ \sqrt{\gamma(u)} \pi \right]}.
\end{equation}
Then, $C(\theta,u) /C(0,0)$ is a correlation function.
\item[(II)]  Let $C$ be defined as 
\begin{equation} \label{emi2}
C(\theta,u) = \exp\left\lbrace \rho(u) \cos(\theta) - 1 \right\rbrace \cos \left\lbrace \rho(u) \sin\theta  \right\rbrace.
\end{equation}
Then, $C(\theta,u) $ is a correlation function.
\end{description}
\end{theorem}
Proof of Theorem \ref{thm:circle1} is deferred to Appendix \ref{app:theory}. Examples of variograms $\gamma(\cdot)$ are reported  in Tables 3 and 5 in the Online Supplement. We give Equations (\ref{eq:time8}), (\ref{eq:time9}), and (\ref{eq:time10}) in Appendix \ref{app:time_ex} as examples from Theorem \ref{thm:circle1}. These models do not decay to zero at $t$ gets large. 
If eventual decay to zero is preferred, then we could construct a model $C(\theta,u) = C_5(\theta,u) C_2(u)$, where $C_2(u)$ decays to zero as a function of time, but this construction did not improve predictive performance for these data.

\section{Methods and Models}\label{sec:modeling}

\subsection{Nearest-Neighbor Gaussian Process}\label{sec:nngp}

We envision a model for the Mexico City data as
\begin{align}
Y(\bs,t) &= \bx(\bs,t)^\top \bbeta + w(\bs,t) + \epsilon(\bs,t), \\
w(\bs,t) &\sim \mathcal{N}(0,C((\bs,t),(\bs',t'))),  \nonumber \\
\epsilon(\bs,t) &\overset{iid}{\sim} \mathcal{N}(0,\tau^2),
 \nonumber
\end{align} 
where $Y(\bs,t)$ is square-root ozone measured at the location-time pair $(\bs,t)$, $\bx(\bs,t)$ are relative humidity and temperature measurements, $\bbeta$ are regression coefficients, $w(\bs,t)$ are space-time random effects, and $\epsilon(\bs,t)$ is Gaussian noise. We found that using square-root ozone reduced correlation between the variance of model residuals and predicted mean. Additionally, modeling on the square-root scale led to better predictive performance than modeling on the original scale \citep[see also][]{sahu2007,berrocal2010}. 

Space-time random effects for point-referenced data are often specified through a functional prior using a Gaussian process (GP) \citep[see, e.g.,][]{rasmussen2006,banerjee2014}. Likelihood computations for hierarchical Gaussian process models require inverting a square matrix that is the same size as the data, making Gaussian process models intractable with even moderate amounts of data (for example, $N > 10,000$). 

Many have addressed this computational bottleneck using either low-rank or sparse matrix methods \citep[see][for a review and comparison of these methods]{heaton2017}. Low-rank methods project the original process onto representative points or knots \citep[see, e.g.,][]{higdon2002,banerjee2008,cressie2008,stein2008}; however, these approaches often perform poorly for prediction as they often over-smooth \citep[see][]{stein2014, heaton2017}. 

Alternatively, sparse methods either induce zeros in the covariance matrix using compactly supported covariance functions \citep[see, e.g.,][]{furrer2006,kaufman2008} or in the precision matrix by assuming conditional independence \citep{vecchia1988,stein2004}. We ultimately favor conditional independence approaches because predictive performance is generally better, and the class of valid covariance models is more expansive.

Sparse precision methods date to \cite{vecchia1988} and were furthered by \cite{stein2004} and \cite{bev} to approximate the likelihood of the full GP model using conditional likelihoods. \citet{gramacy2015} and \cite{datta2016a} extend this work to process modeling. The nearest-neighbor Gaussian process is itself a Gaussian process \citep{datta2016a} and has good predictive performance relative to other fast FP methods \citep[See][]{heaton2017}.

If we take a \emph{parent} Gaussian process over $\mathcal{D} \times \S^1 \times \R$, the nearest-neighbor Gaussian process induces sparsity in the precision matrix of the parent process by assuming conditional independence given neighborhood sets \citep{datta2016a}. Let $\mathcal{S} = \{(\bs_1,t_1), (\bs_2,t_2), ..., (\bs_k,t_k) \}$ of $k$ distinct location-time pairs denote the reference set, where time gives natural ordering and spatial ordering induced for observations at identical times. In our case, we order space by latitude, south to north, and take observed data as the reference set. For model validation, we include held-out location-time pairs in the reference set, making prediction part of model fitting.

We define neighborhood sets $N_\mathcal{S} = \{N(\bs_i,t_i); i = 1,...,k \}$ on the reference set, where $N(\bs_i,t_i)$ consists of $m$ nearest-neighbors of $(\bs_i,t_i)$ chosen from $\{ (\bs_1 , t_1) , (\bs_2 , t_2) , ... , (\bs_{i-1} , t_{i-1} ) \}$. Evidently, if $i \leq m+1$, $N(\bs_i,t_i) = \{(\bs_1,t_1), (\bs_2,t_2), ..., (\bs_{i-1},t_{i-1} ) \}$. Together, $\mathcal{S}$ and $N_\mathcal{S}$ define a Gaussian directed acyclic graph (DAG) with joint distribution
\begin{equation}
p( \bw_\mathcal{S} ) = \prod^k_{i=1} p(\bw(\bs_i,t_i)|\bw_{N(s_i)})
= \prod^k_{i=1} N(\bw(\bs_i,t_i)| \bB_{(\bs_i,t_i)} \bw_{N(\bs_i,t_i)}, \bF_{(\bs_i,t_i)}), 
\end{equation}  
where 
$\bB_{\bs_i,t_i} = C_{(\bs_i,t_i),N(\bs_i,t_i)} C_{N(\bs_i,t_i)}^{-1}$, $\bF_{(\bs_i,t_i)} = C((\bs_i,t_i),(\bs_i,t_i)) - \bB_{(\bs_i,t_i)} C_{N(\bs_i,t_i),(\bs_i,t_i)}$,
and $\bw_{N(\bs_i)}$ is the subset of $\bw_\mathcal{S}$ corresponding to neighbors $N(\bs_i)$ \citep{datta2016a}. Like other GP models, the NNGP can be used for hierarchically spatial random effects.

\subsection{Neighbor Selection}\label{sec:neighbor}

Optimal neighbor selection is not simple, and ``best'' subsets vary depending on covariance function and associated parameters \citep[see, e.g.,][]{vecchia1988,datta2016c}. Defining nearest-neighbors only in terms of distance can lead to models that are uniformly sub-optimal compared to other neighbor selections \citep{stein2004}, but nearest-neighbor approaches are often used for simplicity. \cite{gramacy2015} provide a thorough discussion on how to improve upon nearest-neighbor selections.

For daily monitoring data, \cite{stein2005} argues using most-recent neighbors. In the same setting, \cite{datta2016c} discuss two approaches for selecting neighbors: simple neighbor selection and dynamic neighbor selection. For $m$ neighbors, simple selection involves choosing $\sqrt{m}$ nearest spatial neighbors over the $\sqrt{m}$ most recent times, including current times. For dynamic neighbor selection, neighbors change within model fitting to select the most correlated neighbors. In some nonseparable space-time settings, possible neighbors can be enumerated simply, limiting the computation burden of dynamic neighbor selection \citep{datta2016c}. Enumeration of possible neighbors for periodic time and quasi-periodic time covariance models is intractable because the covariance function has many local maxima over time.  

Even for fixed periods, the pattern of covariance seasonality varies greatly depending on the model and its parameters. Thus, neighborhood selection may be specific to the model and data. Our argument is simple. Including periodic peaks, as well as locations near periodic peaks improves predictive performance. We provide an empirical discussion to this point in the Online Supplement.

\subsection{Priors, Prediction, and Model Comparison}\label{sec:prior}

We begin here by discussing the prior distributions for our model parameters. We use inverse-gamma prior distributions for $\tau^2$ and $\sigma^2$ with 2.1 and 10 for the shape and rate parameters. For square-root ozone concentrations, these selections are not overly informative. For compactly supported covariance parameters, we use uniform prior distributions. For parameters that are strictly positive, we use gamma prior distributions with shape and rate of 0.01. We assume that regression coefficients \emph{a priori} follow independent normal distributions with mean zero and variance $10^3$. For this model formulation, model fitting details can follow either \cite{datta2016a} or the collapsed sampler in \cite{finley2018}.

Predictions at unobserved times and locations can depend on any subset of the reference set $\mathcal{S}$. Because this is a retrospective analysis, we use past and future lags for prediction, where we again consider neighbor selection carefully to account for seasonality. In our case, we use the five nearest spatial neighbors with 1, 2, 23, 24, 25, and 168-hour lags back and forward, as well as simultaneous spatial neighbors. Random effects at unobserved location-time pairs have a conditional normal distribution given by
\begin{equation}
w(\bs,t) \vert \bw_{N(\bs,t)} \sim \mathcal{N}\left( C_{(\bs,t),N(\bs,t)} C_{N(\bs,t)}^{-1} \bw_{N(\bs,t)} , \\
C((\bs,t),(\bs,t)) - C_{(\bs,t),N(\bs,t)} C_{N(\bs,t)}^{-1} C_{(\bs,t),N(\bs,t)}^\top \right). \label{eq:cond}
\end{equation}
Then, the posterior prediction $Y(\bs,t) \mid  \bY$ done using posterior samples of $\bbeta$, $w(\bs,t)$, and $\tau^2$ via composition sample \citep[see, e.g.,][]{gelman2014}. To obtain estimates of $\bx(\bs,t)$ at unmonitored locations, we weight the 24 simultaneously observed covariates proportional to inverse squared distance. Predictions at held-out location-time pairs are used to compare competing models. 

We fit models on the square-root scale but compare models using predictive criteria on the original scale. To validate our space-time model, we hold out all observations (i.e. all stations) at 20\% of the observed hours and compare models using root mean squared prediction error (RMSPE), mean absolute prediction error (MAPE), $100\times (1-\alpha)$ \% prediction interval coverage (CVG), and continuous ranked probability scores (CRPS) \citep{gneiting2007}, where 
\begin{equation} 
\text{CRPS}(F_i,y_i) = \int^\infty_{-\infty} (F_i(x) -  \bone(x \geq y_i) )^2 dx = \bE|Y_i - y_i | - \frac{1}{2}\bE | Y_i - Y_{i'} | .
\end{equation}
We use $M$ posterior predictive samples for a Monte Carlo approximation \citep[see, e.g.,][]{kruger2016}, $$\text{CRPS}(\hat{F}_i,y_i) = \frac{1}{M} \sum_{j=1}^M |Y_j - y_i| - \frac{1}{2M^2} \sum_{j=1}^M \sum_{k=1}^M | Y_j - Y_k|.$$
We then average $\text{CRPS}(\hat{F}_i,y_i)$ over held-out data. Because CRPS considers how well the entire predictive distribution matches the observed data rather than only the predictive mean (MAPE and RMSPE) or quantiles (CVG). Furthermore, CRPS is a proper scoring rule \citep{gneiting2007}; thus, we prefer it as a selection criterion. 

To assess predictions for all observations at a held-out hour jointly, we consider the energy score (ES), a multivariate generalization of CRPS. For a set of multivariate predictions $\bY$, ES is
\begin{equation}\label{eq:es}
\text{ES}(P,\by) = \frac{1}{2} E_P \left\| \bY - \bY' \right\|^\eta - E_P \left\| \bY - \by \right\|^\eta,
\end{equation} 
where $\by$ is an observation, $\eta \in (0,2)$, and $P$ is a probability measure \citep{gneiting2007}. It is common to fix $\eta = 1$ \citep[see, e.g.,][]{gneiting2008,jordan2017}. For a set of $M$ MCMC predictions $ \bY = \bY_1,...,\bY_M$ for a held-out observation $\by$, the empirical ES reduces to 
$$\text{ES}(\bY,\by) = \frac{1}{M} \sum^M_{j=1} \left\| \bY_j - \by \right\|  - \frac{1}{2M^2} \sum^M_{i=1} \sum^M_{j=1} \left\| \bY_i - \bY_j \right\|,$$
as was done in \citet{gneiting2008}. Like CRPS, ES is a proper scoring rule \citep{gneiting2007}. To obtain a single value, average ES over held-out data.

\subsection{Ozone Exposure Respiratory Risk Assessment}\label{sec:risk}

We use the daily respiratory risk model presented in \cite{chiogna2011} to assess risk attributable to ground-level ozone. In their paper, \cite{chiogna2011} model the short-term effects of summer ozone levels on all hospital admissions and respiratory hospital admissions. For this goal, they compare 115 models using cross-validation. Although the risk model of \cite{chiogna2011} does not include space, it can be applied location-wise. 

\cite{chiogna2011} found that the best model included three measures associated with an ozone threshold of $T = 60$ ppb:  number of the hours exceeding $T$ on day $d$ (we call this $H(\bs,d)$), the difference between the maximum daily ozone level and $T$ (max - threshold) $D(\bs,d)$, and the lagged average nighttime ozone from the last three days $O_n(\bs,d)$ (nighttime is defined to be 9 PM and 8 AM). We used fewer days to calculate $O_n(\bs,d)$ if data from the last three nights were not available (e.g. April 2 only has one night to estimate risk). Then, the relative risk of respiratory hospital admissions $r(\bs,d)$ on day $d$ at location $\bs$ is
\begin{equation}
 r(\bs,d) = 0.864 \exp \left( 5.020\times10^{-4} H(\bs,d)D(\bs,d) + 5.714\times 10^{-3}O_n(\bs,d)   \right),
\end{equation}
where $ 5.020\times10^{-4}$ and $5.714\times 10^{-3}$ are regression coefficients for $H(\bs,d)D(\bs,d)$ and $O_n(\bs,d)$, respectively. The scaling factor $0.864$ makes the average risk at the 24 ozone monitoring stations over the year equal to one. The scale of the coefficients differs from those presented in \cite{chiogna2011} because we use ozone levels in ppb instead of $\mu g/ m^3$.

By adopting this model, we make many assumptions that we outline here. First, we assume that the effects of ozone in Milano, Italy are like those in Mexico City, Mexico. While the marginal effect may be the same, it is likely that the effects of ozone are influenced by other factors like other pollutants and weather. Second, we adopt their choice of threshold where ozone begins to be harmful (they consider three thresholds), although several question using thresholds altogether \citep[See, e.g.,][]{kim2004}. Additionally, we consider the difference between the maximum daily level compared to a threshold instead of integrated ozone as was used in \cite{arisido2016}. Noting these limitations, this model enables us to estimate respiratory risk on a daily level using predicted hourly ozone levels.

\section{Results and Discussions}\label{sec:space_time}

To select a covariance model, we hold out all observations at 20\% of hours in April and May, as discussed in Section \ref{sec:modeling}. This hold-out approach allows us to score models using the energy score, as well as univariate criteria like CRPS, RMSPE, MAPE, and 90\% prediction interval coverage. While RMSPE and prediction interval coverage are the most common criteria, we rely upon CRPS and ES most because they are proper scoring rules \citep{gneiting2007} and compare the whole predictive distribution to held-out values. Out-of-sample predictions are made using 25,000 posterior samples after a burn-in of 5,000 iterations.

We present results for a variety of covariance models on $\mathcal{D} \times \S^1 \times \R$ or a subset of this space. These models are given in Table \ref{tab:space_time_comp}. We first consider a completely separable covariance model on $\mathcal{D} \times \S^1 \times \R$, Model 1, which is used to motivate the use of nonseparable models. We then consider two examples from \cite{gneiting2002}. The first, given as Model 2, ignores periodicity. We also consider this same covariance model multiplied by an exponential covariance function that takes $\theta$ as an argument (Model 3), an example of construction (\bB) in Section \ref{sec:covar}. Then, we consider two examples based on the class of nonseparable covariance models presented by \cite{shirota2017} (Models 4 and 5), where Model 4 ignores temporal decay and Model 5 is an example of construction (\bA). Then, we give two examples of nonseparable circular cross linear time covariance models, one from \cite{white2018b} and the other from Theorem \ref{thm:circle1}, where both models are multiplied by an exponential spatial covariance function (Models 6 and 7). Both Models 6 and 7 are examples of construction (\bC). Other models from these classes were considered; however, these represent the models with the best predictive performance. 

All models except Model 4 use the five nearest spatial neighbors (including the location itself) with 1, 2, 23, 24, 25, and 168-hour lags. We use neighbors at the same time as the observation but exclude the observation itself. Thus, we condition on at most 34 neighbors. For Model 4, we exclude lags 24 and 168 at the location itself to maintain positive-definite covariance matrices; instead, we include lags 3, 167, and 169 to compensate. The results of the model comparison are given in Table \ref{tab:space_time_comp}. 

The results highlight the importance of modeling the entire space ($\mathcal{D} \times \S^1 \times \R$) and utilizing nonseparable models. Models that excluded one of these subspaces (Models 2 and 4) had the worst predictive performance. The completely separable model (Model 1) was next worse in prediction, motivating the need for nonseparable models. Each of the constructive approaches, denoted (\bA), (\bB), and (\bC), proposed in Section \ref{sec:covar} performed well in prediction. Ultimately, Model 7, derived from Theorem \ref{thm:circle1} and using construction (\bC), performed best in terms of ES, CRPS, MAPE, and RMSPE for these data. Thus, we use this model for prediction at unmonitored locations.

Explicitly, the final covariance model is 
\begin{equation}\label{eq:final_mod}
C(h, \theta,u) = \exp\left\lbrace \exp\left[ - \left( \frac{|u|}{c_t}\right)^{\alpha} \right] \cos(\theta) - \frac{ h}{c_s} - 1\right\rbrace \cos\left \lbrace \exp\left[ - \left( \frac{|u|}{c_t}\right)^{\alpha} \right]\sin(\theta) \right \rbrace,
\end{equation}
where $(h, \theta, u) \in [0,\infty) \times [0,\pi] \times [0,\infty)$, $c_t, c_s >0$ and $\alpha \in (0,2]$. The parameter $c_t$ governs the temporal range, $c_s$ is the spatial range parameter, and $\alpha$ is a smoothness parameter.  The parameter $c_s$ is most accurately understood as the spatial range for simultaneous observations, and $c_t$ is the temporal range at a fixed location. One benefit of this model is that it has few parameters compared with many of its competitors.

Posterior summaries for regression coefficients and covariance parameters are given in Table \ref{tab:summaries}. The regression parameters agree with our hypotheses that relative humidity is negatively related to ozone levels and the temperature is positively associated with ozone. The mean of the spatial range parameter $c_s$ is 22.32 km. The posterior mean of the temporal range parameter is 86.90 hours, meaning that correlation for this model persists over several days. However, these range parameters have limited interpretability, as discussed above.

We limit our predictions to the convex hull of the monitoring network to not spatially extrapolate. Using these posterior predictions in April and May, we can assess probabilities of exceeding Mexican ambient air quality standards and examine respiratory risk rates during Mexico City's peak ground-level ozone season. 

\begin{figure}[H]
\begin{center}
\includegraphics[width=\textwidth]{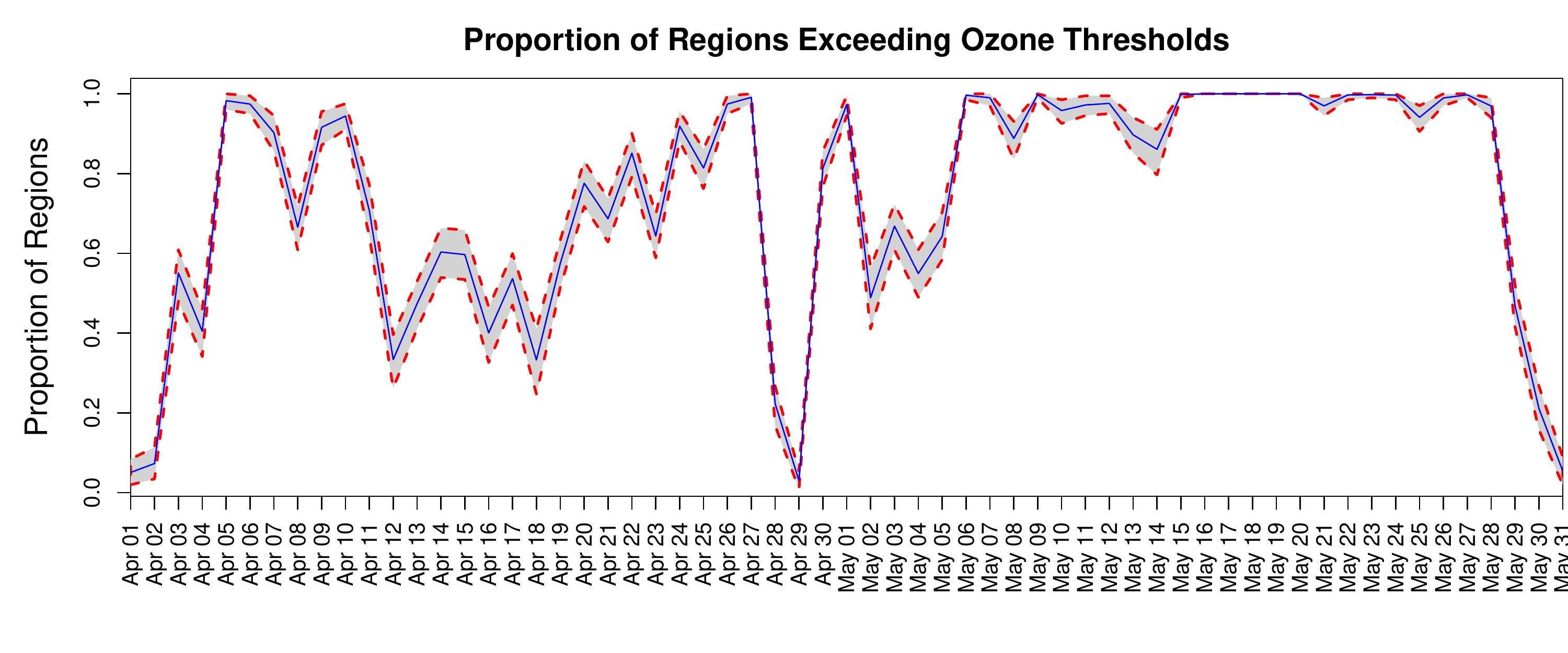}
      \caption{Estimated proportion of the city that exceeds either one-hour and eight-hour average ozone limits at least once that day.}\label{fig:time_exceed}
\end{center}
\end{figure}

To summarize exceedance results temporally, we plot posterior means and 95\% credible intervals for the estimated proportion of the city that exceeds national ozone standards at least once on that day in Figure \ref{fig:time_exceed}. In April, there are several peaks in exceedance probability; however, from May 6 to May 28, the proportion of the city with at least one daily ozone exceedance is near one. 

\begin{figure}[H]
\begin{center}
   \begin{subfigure}[b]{.32\textwidth}
\includegraphics[width=\textwidth]{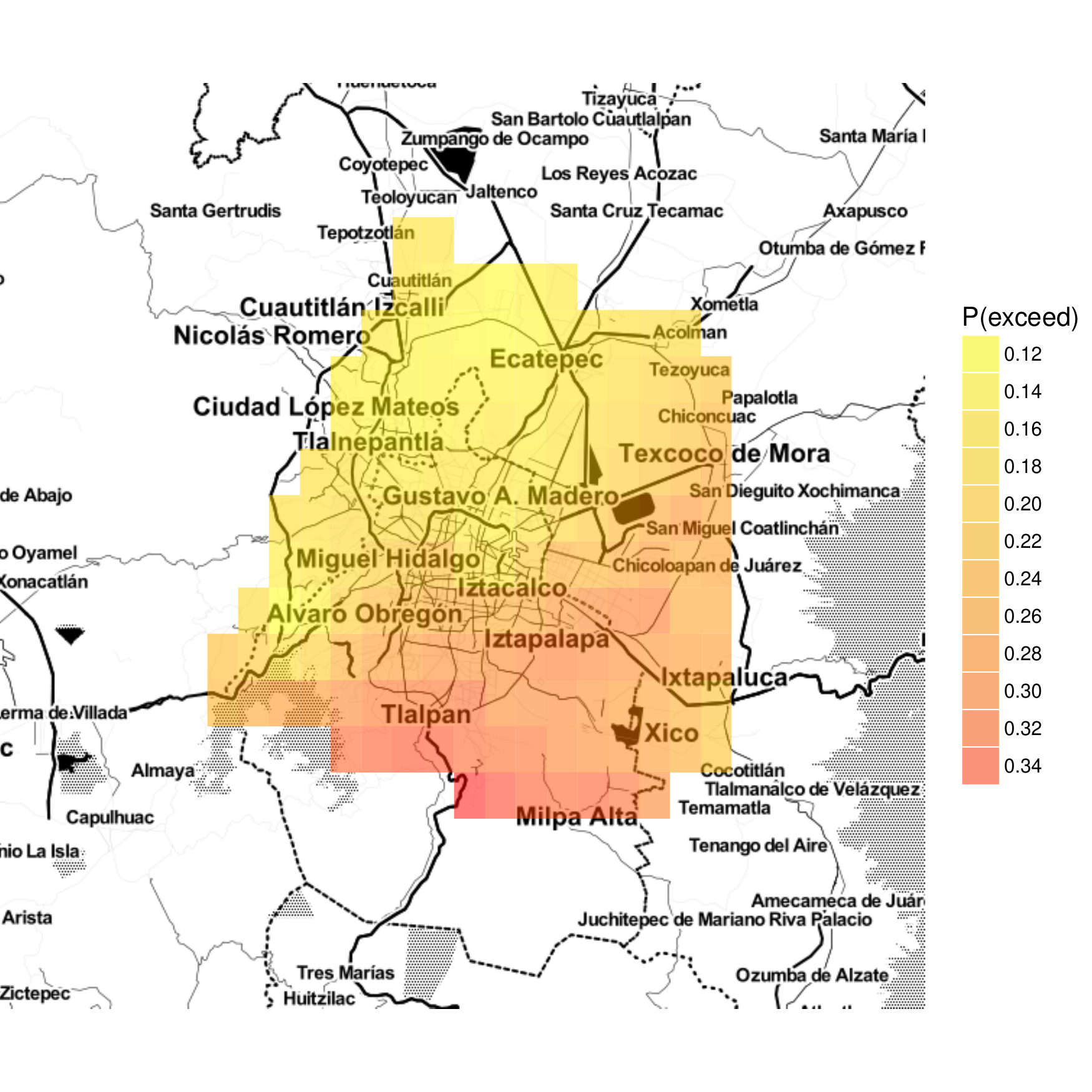}
      \subcaption{Exceedance probability}\label{fig:exceed_spat}
   \end{subfigure}
   \begin{subfigure}[b]{.32\textwidth}
\includegraphics[width=\textwidth]{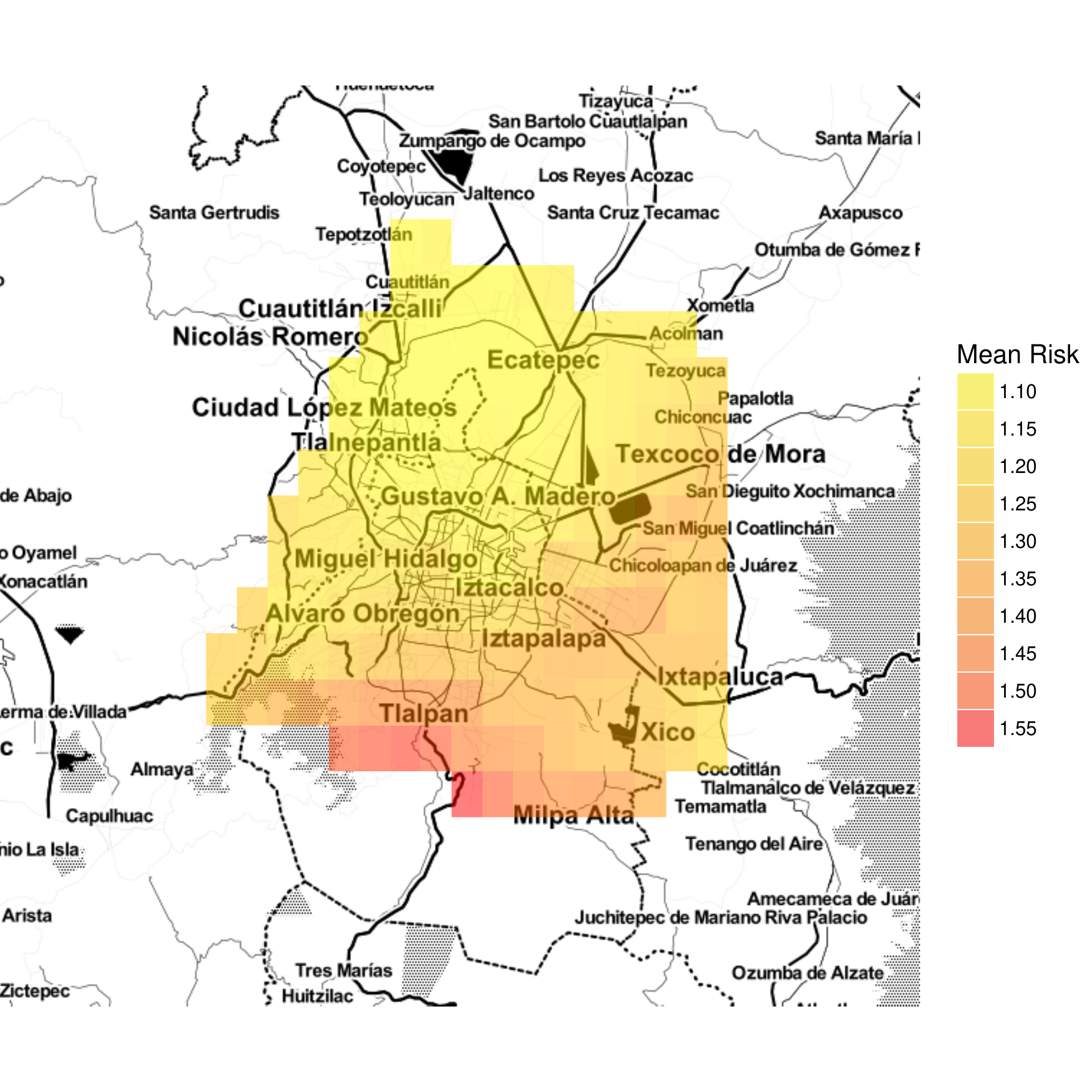}
      \subcaption{Mean risk}\label{fig:mean_risk}
   \end{subfigure}
   \begin{subfigure}[b]{.32\textwidth}
\includegraphics[width=\textwidth]{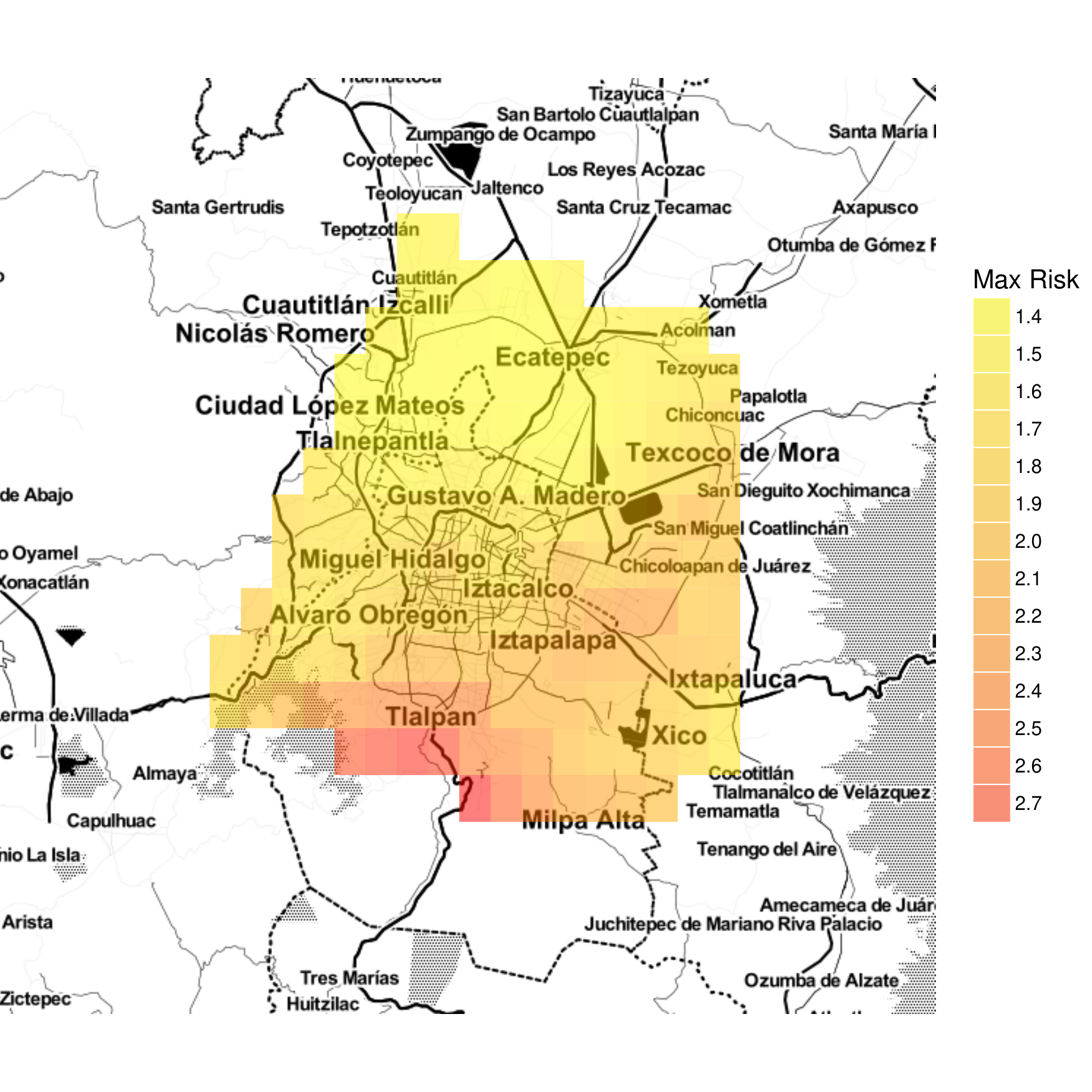}
      \subcaption{Maximum risk}\label{fig:max_risk}
   \end{subfigure}
\caption{Spatial summaries of (Left) exceedance probability and (Center and Right) respiratory risk for ozone. The exceedance probability considers both one-hour and eight-hour average ozone limits. }\label{fig:spat_plots}
\end{center}
\end{figure}

For each predictive location, we plot the estimated proportion of hours when Mexican ground-level ozone standards (either one-hour or eight-hour ozone) are exceeded in Figure \ref{fig:exceed_spat}. This follows the pattern in Figure \ref{fig:station_loc}, but the degree of variation between locations is striking. We estimate that locations in southern Mexico City exceed national ozone regulations nearly three times as often as north, west, and central Mexico City. 

In Figures \ref{fig:mean_risk}, we give the mean relative risk averaged over all days in April and May, and we plot the estimated maximum relative risk over the same time period in Figure \ref{fig:max_risk}. Although the spatial patterns of the means and maxima are similar, the scale of these plots differs significantly. In terms of mean risk, the most extreme regions have respiratory risks 50\% higher than the least extreme regions. In contrast, regions of the highest maximum risk in the peak season are nearly twice that of regions with the lowest maximum risk. These plots demonstrate the degree of increased health risk determined solely by where one lives or works.

\begin{figure}[H]
\begin{center}
\includegraphics[width=\textwidth]{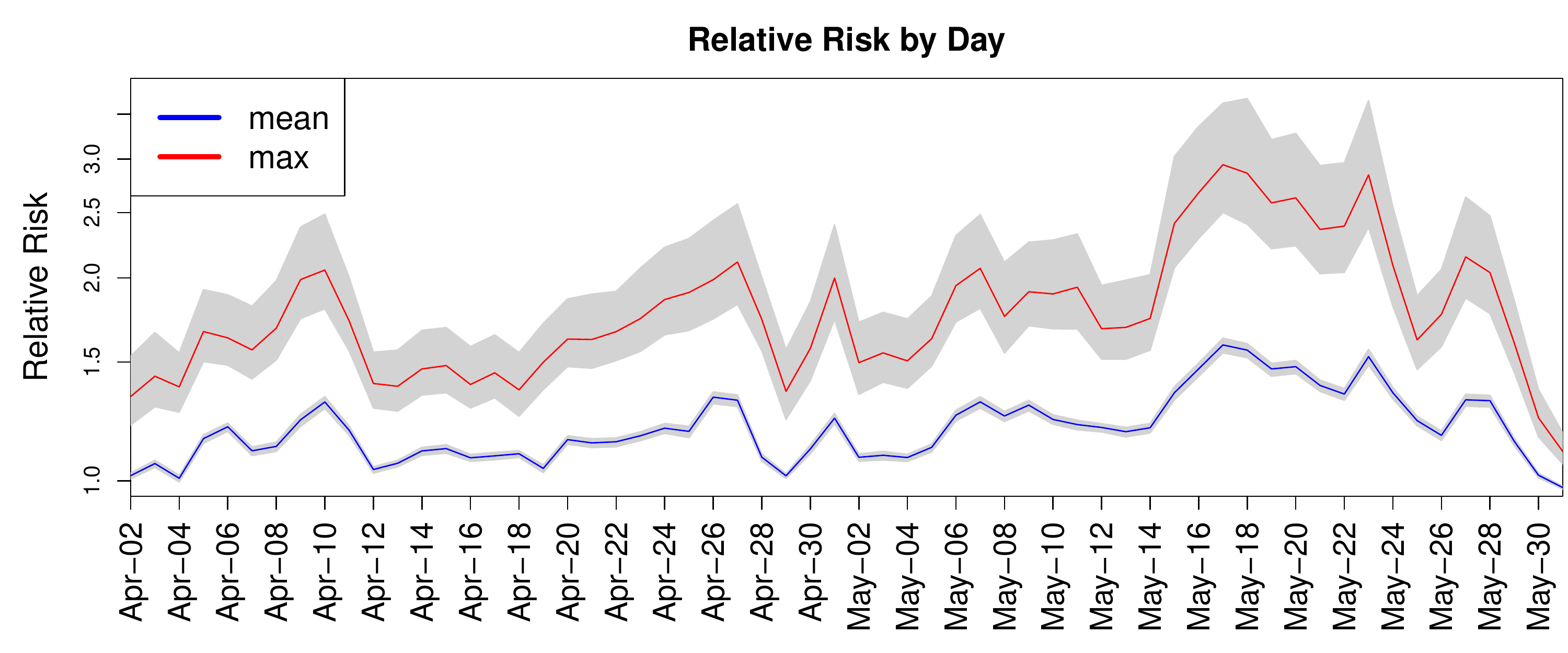}
\caption{Estimated mean and maximum respiratory risk over the city.}\label{fig:time_risk}
\end{center}
\end{figure}

The estimated mean and maximum respiratory health risk over Mexico City (with 95\% credible intervals) are given as a function of the day in Figure \ref{fig:time_risk}. Ozone risk peaks May 17 to May 23 in terms of both the mean and maximum, but the changes in the maximum risk are much more drastic than those in mean risk. These maxima correspond to extreme ozone levels in south Mexico City where the estimated risk is 2.7 times the annual average and nearly two times the mean risk on the same day.

These results highlight how extreme the space-time variability in both exceedance probability and respiratory health risk is within Mexico City during its peak ozone season. According to the risk model from \cite{chiogna2011}, those living in regions with extreme ozone levels are about 50\% more likely to be admitted to the hospital due to ozone exposure compared to those in less polluted areas. Although this is not a component of the risk model, increased respiratory risk impacts at-risk populations (e.g. the elderly) at a much higher rate than healthier sub-populations \citep[see, e.g.,][]{bell2004}.

\section{Discussion and Conclusions}\label{sec:conc}

We have discussed the monitoring network for ozone and analyzed these data in April and May of 2017. For these data, we develop new classes of covariance models that account for daily seasonality. We apply these covariance models to the Mexico City ozone data using nearest-neighbor Gaussian processes and select a model using predictive criteria on a hold-out dataset. This model is then used for prediction to assess compliance with Mexican ambient air quality standards and respiratory risk due to ozone exposure. In this analysis, we identify regions and times where exceedance probabilities and ozone risk factors are particularly high.  

An interesting follow-up study would account for people's movement over the day, giving individualized risk assessments. In addition to this, we could combine our modeling approaches with hospital admission data in Mexico City, allowing us to establish more accurate respiratory risk measures for Mexico City, rather of relying on other methods. 

Further theoretical work could address how the constraints on the functions $\varphi$ and $\psi$ in \cite{shirota2017} can be relaxed while preserving positive-definiteness. Here, we have considered models that are partially nonseparable over $\mathcal{D} \times \S^1 \times \R$ but not over all three spaces. Thus, further theoretical work to establish such classes is justified.

\appendix

\section{Proof of Theorem \ref{thm:circle1}}\label{app:theory}
We start by noting that arguments in \cite{berg-porcu} show that the functions $C$ defined at points {\bf (I)} and {\bf (II)} are positive definite if and only if they can both be written as 
\begin{equation}
\label{berg-porcu-expansion} 
 C(\theta,u) = \sum_{k=0}^{\infty} b_{k}(u) \cos (k \theta), \qquad (\theta,u) \in [0,\pi] \times \R, 
\end{equation}
where uniquely determined sequence of positive definite functions $\{ b_{k}(\cdot)\}_{k=0}^{\infty}$ has $\sum_k b_{k}(0) < \infty$. To show {\bf (I)}, we resort to $[1.445.2]$ in \cite{grad}:
$$ \sum_{k=1}^{\infty} \frac{\cos(k x)}{(k^2+a^2)} = \frac{\pi}{2} \frac{\sinh \left ( a(\pi-x)\right ) }{\sinh(\pi a )}, \qquad a >0, x \in \R. $$
The function in Equation (\ref{emi1}) admits an expansion of the type (\ref{berg-porcu-expansion}) with coefficients $b_{k}(u)= (k^2 + \gamma(u))^{-1}$ for $k=0,1,\ldots$. It can be namely verified that $\sum_k b_k(0)$ is finite. Thus, the proof is completed. 
As for Assertion {\bf (II)}, Equation (\ref{emi2}) comes straight by considering the expansion in [1.449.2] in \cite{grad}:
$$ \sum_{k=0}^{\infty} \frac{p^k \cos (kx)}{k!} = {\rm e}^{p \cos (x)} \cos (p \sin x), $$
which is absolutely convergent provided $p^2 \le 1$. We now replace $p$ with the correlation function $\rho(\cdot)$ to obtain (\ref{emi2}). The proof is completed. \hfill $\Box$

\section{Covariance Examples}\label{app:examples}

\subsection{Temporal Modeling}\label{app:time_ex}

We begin with two purely circular covariance examples from \cite{gneiting2013} in (\ref{eq:time2}) and (\ref{eq:time3}). For $\alpha \in (0,1]$ and $\lambda,c_p >0$, we define
 \begin{align} 
 C(\theta) &= \exp\left[ - \left(\theta / c_p \right)^\alpha \right] \qquad \theta \in [0,\pi]\qquad \text{and }\label{eq:time2} \\
 C(\theta) &=  \left[ 1 +  \left( \theta / c_p \right)^\alpha \right]^{-\lambda} \qquad \theta \in [0,\pi]. \label{eq:time3}
 \end{align}

In addition to the Mat\'ern model presented in Section \ref{sec:covar}, we consider a pair of covariance models on $\S^1 \times \R$ from what we call the inverted Gneiting class in \citep{porcu2016}. Let $\delta,\lambda>0$, and $\alpha, \beta, \gamma \in (0,.1]$. Also, let $c_s$ and $c_t$ be positive parameters. Then
\begin{equation}\label{eq:time4}
C(\theta,u) = \frac{\sigma^2}{\left( 1 + \left(\frac{\theta}{c_s}\right)^\alpha \right)^{\delta + \beta/2} } \exp\left( - \frac{ | u| ^{2\gamma} }{c_t^{2 \gamma} \left( 1 + \left(\frac{\theta}{c_s}\right)^\alpha \right)^{\beta \gamma} } \right), \qquad (\theta,u) \in [0,\pi] \times \R,
\end{equation}
where $\delta>0$, and where  $\beta, \alpha$ and $\gamma$ belong to the interval $(0,1]$. The second we consider uses the generalized Cauchy covariance function \citep[][see Tables 4 and 5 in the Online Supplement]{GS2004} for the temporal margin,
\begin{equation}\label{eq:time5}
C(\theta,u) = \frac{\sigma^2}{\left( 1 + \left(\frac{\theta}{c_s}\right)^\alpha \right)^{\delta + \beta/2} } \left( 1+ \frac{ |u| ^{2\gamma} }{c_t^{2\gamma} \left( 1 + \left(\frac{\theta}{c_s}\right)^\alpha \right)^{\beta \gamma} } \right)^{-\lambda},  \qquad (\theta,u) \in [0,\pi] \times \R.
\end{equation}

Examples from \cite{white2018b} include
\begin{equation}\label{eq:time6}
C(\theta,u) = \frac{\sigma^2}{\left( 1 + \left( \frac{\mid u\mid }{c_t} \right)^\alpha \right)^{\delta + \beta d/2} } \exp\left( - \frac{ \theta^{\gamma} }{c_{s}^{ \gamma}\left( 1 + \left(\frac{\mid u\mid }{c_t}\right)^\alpha \right)^{\beta \gamma} } \right),  \qquad (\theta,u) \in [0,\pi] \times \R,
\end{equation}
with $\delta >0$, $\beta, \gamma \in (0,1]$ and $\alpha \in (0,2]$. For the second, we propose again a generalized Cauchy covariance function for the spatial margin, obtaining
\begin{equation}\label{eq:time7}
C(\theta,u) = \frac{\sigma^2}{\left( 1 + \left(\frac{\mid u\mid }{c_t}\right)^\alpha \right)^{\delta + \beta d/2} } \left( 1+ \frac{ \theta^{\gamma} }{c_s^{\gamma}  \left( 1 + \left(\frac{\mid u\mid }{c_t}\right)^\alpha \right)^{\beta \gamma} } \right)^{-\lambda},  \qquad (\theta,u) \in [0,\pi] \times \R,
\end{equation}
where  $\delta >0$, $\beta, \gamma \in (0,1]$, $\alpha \in (0,2]$ and $\lambda >0$.

From Theorems \ref{thm:circle1}, we give three examples. Let $\alpha \in (0,2]$ and $c_t,\lambda > 0$. Then, we propose
{\footnotesize
\begin{align}
C(\theta,u) &=  \left[1 + \left( \frac{u}{c_t}\right)^\alpha \right]^{-\beta}   + \frac{\pi}{2} \frac{\sinh \left(  (\pi - \theta) \sqrt{\left[1 + \left( \frac{u}{c_t}\right)^\alpha \right]^{\beta} }  \right) }{\sinh \left( \pi \sqrt{\left[1 + \left( \frac{u}{c_t}\right)^\alpha \right]^{\beta}}  \right)} \qquad &(\theta,u) \in [0,\pi] \times \R, \label{eq:time8} \\
C(\theta,u) &= \exp\left\lbrace \left[ 1 + \left( \frac{u}{c_t} \right)^\alpha \right]^{-\lambda} \cos(\theta) - 1\right\rbrace \cos\left \lbrace \left[ 1 + \left( \frac{u}{c_t} \right)^\alpha \right]^{-\lambda} \sin(\theta) \right \rbrace \qquad &(\theta,u) \in [0,\pi] \times \R, \label{eq:time9} \\
C(\theta,u) &= \exp\left\lbrace \exp\left[ - \left( \frac{u}{c_t}\right)^{\alpha} \right] \cos(\theta) - 1\right\rbrace \cos\left \lbrace \exp\left[ - \left( \frac{u}{c_t}\right)^{\alpha} \right]\sin(\theta) \right \rbrace \qquad &(\theta,u) \in [0,\pi] \times \R,\label{eq:time10}
\end{align}
}

\subsection{Space-Time Models}\label{app:space_time_ex}

To create effective space-time models, we suggest combining models from Appendix \ref{app:time_ex} with valid spatial models. Additionally, we provide examples of covariance models on $\mathcal{D} \times \R$ and $\mathcal{D} \times \S^1$ which can be combined with valid models on $S^1$ and $\R$ given in Appendix \ref{app:time_ex}.

First, we look at nonseparable covariance models on ${\cal D} \times \R$ from the so-called Gneiting class from \cite{gneiting2002} given in (\ref{gneiting_form}). From this class, we give two examples.
For the first, we propose again a generalized Cauchy covariance function for the spatial margin, obtaining
\begin{equation}\label{eq:space2}
C(h,u) = \frac{\sigma^2}{\left( 1 + \left(\frac{u }{c_t}\right)^\alpha \right)^{\delta + \beta d/2} } \left( 1+ \frac{ h^{2\gamma} }{c_s^{2\gamma}  \left( 1 + \left(\frac{u }{c_t}\right)^\alpha \right)^{\beta \gamma} } \right)^{-\lambda},  \qquad (h,u) \in [0,\infty)^2,
\end{equation}
for $\delta,\lambda >0$, $\beta, \gamma \in (0,1]$ and $\alpha \in (0,2]$. To incorporate periodicity, giving a model on ${\cal D}  \times \S^1 \times \R$, one take the product of one of these models with, e.g., (\ref{matern}) or (\ref{eq:time2}).

\cite{shirota2017} extends these models to $\R^2 \times \S^1 $, replacing the temporal lag $u$ with the angular difference on a circle. We give one example in Equation (\ref{eq:space4}).
For the second, we propose again a generalized Cauchy covariance function for the spatial margin, obtaining
\begin{equation}\label{eq:space4}
C(h,\theta) = \frac{\sigma^2}{\left( 1 + \left(\frac{ \theta }{c_t}\right)^\alpha \right)^{\delta + \beta d/2} } \left( 1+ \frac{ h^{2\gamma} }{c_s^{2\gamma}  \left( 1 + \left(\frac{ \theta }{c_t}\right)^\alpha \right)^{\beta \gamma} } \right)^{-\lambda},  \qquad \qquad (h,\theta) \in [0,\infty) \times [0,\pi],
\end{equation}
for $\delta,\lambda >0$, $\beta, \gamma \in (0,1]$ and $\alpha \in (0,2]$. To incorporate a decaying covariance over linear time, one simply takes the product of one of these models with a covariance model on $\R$.

\bibliographystyle{apalike}
\bibliography{refs_mexico}

\begin{thebibliography}{}

\bibitem[Arisido, 2016]{arisido2016}
Arisido, M.~W. (2016).
\newblock Functional {M}easure of {O}zone {E}xposure to {M}odel {S}hort-term
  {H}ealth {E}ffects.
\newblock {\em Environmetrics}, 27(5):306--317.

\bibitem[Banerjee et~al., 2014]{banerjee2014}
Banerjee, S., Carlin, B.~P., and Gelfand, A.~E. (2014).
\newblock {\em Hierarchical {M}odeling and {A}nalysis for {S}patial {D}ata}.
\newblock CRC Press.

\bibitem[Banerjee et~al., 2008]{banerjee2008}
Banerjee, S., Gelfand, A.~E., Finley, A.~O., and Sang, H. (2008).
\newblock Gaussian {P}redictive {P}rocess {M}odels for {L}arge {S}patial {D}ata
  {S}ets.
\newblock {\em Journal of the Royal Statistical Society: Series B (Statistical
  Methodology)}, 70(4):825--848.

\bibitem[Barraza-Villarreal et~al., 2008]{barraza2008}
Barraza-Villarreal, A., Sunyer, J., Hernandez-Cadena, L., Escamilla-Nu{\~n}ez,
  M.~C., Sienra-Monge, J.~J., Ram{\'\i}rez-Aguilar, M., Cortez-Lugo, M.,
  Holguin, F., Diaz-S{\'a}nchez, D., and Olin, A.~C. (2008).
\newblock Air {P}ollution, {A}irway {I}nflammation, and {L}ung {F}unction in a
  {C}ohort {S}tudy of {M}exico {C}ity {S}choolchildren.
\newblock {\em Environmental Health Perspectives}, 116(6):832.

\bibitem[Bell et~al., 2004]{bell2004}
Bell, M.~L., McDermott, A., Zeger, S.~L., Samet, J.~M., and Dominici, F.
  (2004).
\newblock Ozone and {S}hort-term {M}ortality in 95 {US} {U}rban {C}ommunities,
  1987-2000.
\newblock {\em Journal of the American Medical Association},
  292(19):2372--2378.

\bibitem[Bell et~al., 2006]{bell2006}
Bell, M.~L., Peng, R.~D., and Dominici, F. (2006).
\newblock The {E}xposure--{R}esponse {C}urve for {O}zone and {R}isk of
  {M}ortality and the {A}dequacy of {C}urrent {O}zone {R}egulations.
\newblock {\em Environmental Health Perspectives}, 114(4):532.

\bibitem[Berg and Porcu, 2017]{berg-porcu}
Berg, C. and Porcu, E. (2017).
\newblock From {S}choenberg {C}oefficients to {S}choenberg {F}unctions.
\newblock {\em Constructive Approximation}, 45(2):217--241.

\bibitem[Berrocal et~al., 2010]{berrocal2010}
Berrocal, V.~J., Gelfand, A.~E., and Holland, D.~M. (2010).
\newblock A {S}patio-{T}emporal {D}ownscaler for {O}utput from {N}umerical
  {M}odels.
\newblock {\em Journal of Agricultural, Biological, and Environmental
  Statistics}, 15(2):176--197.

\bibitem[Bevilacqua et~al., 2012]{bev}
Bevilacqua, M., Gaetan, C., Mateu, J., and Porcu, E. (2012).
\newblock Estimating {S}pace and {S}pace-{T}ime {C}ovariance {F}unctions for
  {L}arge {D}ata {S}ets: {A} {W}eighted {C}omposite {L}ikelihood {A}pproach.
\newblock {\em Journal of the American Statistical Association},
  107(497):268--280.

\bibitem[Bravo-Alvarez and Torres-Jard{\'o}n, 2002]{bravo2002}
Bravo-Alvarez, H. and Torres-Jard{\'o}n, R. (2002).
\newblock {A}ir {P}ollution {L}evels and {T}rends in the {M}exico {C}ity
  {M}etropolitan {A}rea.
\newblock In {\em Urban Air Pollution and Forests}, pages 121--159. Springer.

\bibitem[Chiogna and Pauli, 2011]{chiogna2011}
Chiogna, M. and Pauli, F. (2011).
\newblock Modelling {S}hort-term {E}ffects of {O}zone on {M}orbidity: {A}n
  {A}pplication to the {C}ty of {M}ilano, {I}taly, 1995--2003.
\newblock {\em Environmental and Ecological Statistics}, 18(1):169--184.

\bibitem[Cressie and Johannesson, 2008]{cressie2008}
Cressie, N. and Johannesson, G. (2008).
\newblock {F}ixed {R}ank {K}riging for {V}ery {L}arge {S}patial {D}ata {S}ets.
\newblock {\em Journal of the Royal Statistical Society: Series B (Statistical
  Methodology)}, 70(1):209--226.

\bibitem[Datta et~al., 2016a]{datta2016a}
Datta, A., Banerjee, S., Finley, A.~O., and Gelfand, A.~E. (2016a).
\newblock {H}ierarchical {N}earest-{N}eighbor {G}aussian {P}rocess {M}odels for
  {L}arge {G}eostatistical {D}atasets.
\newblock {\em Journal of the American Statistical Association},
  111(514):800--812.

\bibitem[Datta et~al., 2016b]{datta2016c}
Datta, A., Banerjee, S., Finley, A.~O., Hamm, N.~A., and Schaap, M. (2016b).
\newblock Nonseparable {D}ynamic {N}earest {N}eighbor {G}aussian {P}rocess
  {M}odels for {L}arge {S}patio-{T}emporal {D}ata with an {A}pplication to
  {P}articulate {M}atter {A}nalysis.
\newblock {\em The Annals of Applied Statistics}, 10(3):1286--1316.

\bibitem[{Diario Oficial de la Federaci{\'{o}}n}, 2014]{nom14b}
{Diario Oficial de la Federaci{\'{o}}n} (2014).
\newblock {Norma Oficial Mexicana NOM-025-SSA1-2014}.

\bibitem[Finley et~al., 2018]{finley2018}
Finley, A.~O., Datta, A., Cook, B.~C., Morton, D.~C., Andersen, H.~E., and
  Banerjee, S. (2018).
\newblock {E}fficient {A}lgorithms for {B}ayesian {N}earest {N}eighbor
  {G}aussian {P}rocesses.
\newblock {\em arXiv preprint arXiv:1702.00434}.

\bibitem[Furrer et~al., 2006]{furrer2006}
Furrer, R., Genton, M.~G., and Nychka, D. (2006).
\newblock {C}ovariance {T}apering for {I}nterpolation of {L}arge {S}patial
  {D}atasets.
\newblock {\em Journal of Computational and Graphical Statistics},
  15(3):502--523.

\bibitem[Gelman et~al., 2014]{gelman2014}
Gelman, A., Carlin, J.~B., Stern, H.~S., Dunson, D.~B., Vehtari, A., and Rubin,
  D.~B. (2014).
\newblock {\em {B}ayesian {D}ata {A}nalysis}, volume~2.
\newblock CRC press.

\bibitem[Gneiting, 1998]{gneiting1998}
Gneiting, T. (1998).
\newblock {S}imple {T}ests for the {V}alidity of {C}orrelation {F}unction
  {M}odels on the {C}ircle.
\newblock {\em Statistics \& Probability Letters}, 39(2):119--122.

\bibitem[Gneiting, 2002]{gneiting2002}
Gneiting, T. (2002).
\newblock {N}onseparable, {S}tationary {C}ovariance {F}unctions for
  {S}pace--{T}ime {D}ata.
\newblock {\em Journal of the American Statistical Association},
  97(458):590--600.

\bibitem[Gneiting, 2013]{gneiting2013}
Gneiting, T. (2013).
\newblock {S}trictly and {N}on-{S}trictly {P}ositive {D}efinite {F}unctions on
  {S}pheres.
\newblock {\em Bernoulli}, 19(4):1327--1349.

\bibitem[Gneiting and Raftery, 2007]{gneiting2007}
Gneiting, T. and Raftery, A.~E. (2007).
\newblock {S}trictly {P}roper {S}coring {R}ules, {P}rediction, and
  {E}stimation.
\newblock {\em Journal of the American Statistical Association},
  102(477):359--378.

\bibitem[Gneiting and Schlather, 2004]{GS2004}
Gneiting, T. and Schlather, M. (2004).
\newblock {S}tochastic {M}odels {T}hat {S}eparate {F}ractal {D}imension and the
  {H}urst {E}ffect.
\newblock {\em SIAM Rev.}, 46:269--282.

\bibitem[Gneiting et~al., 2008]{gneiting2008}
Gneiting, T., Stanberry, L.~I., Grimit, E.~P., Held, L., and Johnson, N.~A.
  (2008).
\newblock {A}ssessing {P}robabilistic {F}orecasts of {M}ultivariate
  {Q}uantities, with an {A}pplication to {E}nsemble {P}redictions of {S}urface
  {W}inds.
\newblock {\em Test}, 17(2):211.

\bibitem[Gradshteyn and Ryzhik, 2007]{grad}
Gradshteyn, I.~S. and Ryzhik, I.~M. (2007).
\newblock {\em Tables of {I}ntegrals, {S}eries, and {P}roducts}.
\newblock Academic Press, Amsterdam, seventh edition.

\bibitem[Gramacy and Apley, 2015]{gramacy2015}
Gramacy, R.~B. and Apley, D.~W. (2015).
\newblock {L}ocal {G}aussian {P}rocess {A}pproximation for {L}arge {C}omputer
  {E}xperiments.
\newblock {\em Journal of Computational and Graphical Statistics},
  24(2):561--578.

\bibitem[Heaton et~al., 2018]{heaton2017}
Heaton, M.~J., Datta, A., Finley, A., Furrer, R., Guhaniyogi, R., Gerber, F.,
  Gramacy, R.~B., Hammerling, D., Katzfuss, M., and Lindgren, F. (2018).
\newblock {M}ethods for {A}nalyzing {L}arge {S}patial {D}ata: {A} {R}eview and
  {C}omparison.
\newblock {\em arXiv preprint arXiv:1710.05013}.

\bibitem[Hern{\'a}ndez-Gardu{\~n}o et~al., 1997]{hernandez1997}
Hern{\'a}ndez-Gardu{\~n}o, E., P{\'e}rez-Neria, J., Paccagnella, A.~M.,
  Pi{\~n}a-Garc{\'\i}a, M.~A., Mungu{\'\i}a-Castro, M.,
  Catal{\'a}n-V{\'a}zquez, M., and Rojas-Ramos, M. (1997).
\newblock {A}ir {P}ollution and {R}espiratory {H}ealth in {M}exico {C}ity.
\newblock {\em Journal of Occupational and Environmental Medicine},
  39(4):299--307.

\bibitem[Higdon, 2002]{higdon2002}
Higdon, D. (2002).
\newblock Space and {S}pace-{T}ime {M}odeling {U}sing {P}rocess {C}onvolutions.
\newblock In {\em Quantitative Methods for Current Environmental Issues}, pages
  37--56. Springer.

\bibitem[Huang et~al., 2018]{huang2018}
Huang, G., Lee, D., and Scott, E.~M. (2018).
\newblock {M}ultivariate {S}pace-{T}ime {M}odelling of {M}ultiple {A}ir
  {P}ollutants and {T}heir {H}ealth {E}ffects {A}ccounting for {E}xposure
  {U}ncertainty.
\newblock {\em Statistics in Medicine}, 37(7):1134--1148.

\bibitem[Huerta et~al., 2004]{huerta2004}
Huerta, G., Sans{\'o}, B., and Stroud, J.~R. (2004).
\newblock A spatiotemporal model for mexico city ozone levels.
\newblock {\em Journal of the Royal Statistical Society: Series C (Applied
  Statistics)}, 53(2):231--248.

\bibitem[Jordan et~al., 2017]{jordan2017}
Jordan, A., Kr{\"u}ger, F., and Lerch, S. (2017).
\newblock {E}valuating {P}robabilistic {F}orecasts with the {R} {P}ackage
  {scoringRules}.
\newblock {\em arXiv preprint arXiv:1709.04743}.

\bibitem[Kahle and Wickham, 2013]{kahle2013}
Kahle, D. and Wickham, H. (2013).
\newblock ggmap: {S}patial {V}iualization with {ggplot2}.
\newblock {\em The R Journal}, 5(1):144--161.

\bibitem[Katzfuss and Guinness, 2017]{katzfuss2017}
Katzfuss, M. and Guinness, J. (2017).
\newblock A {G}eneral {F}ramework for {V}ecchia {A}pproximations of {G}aussian
  {P}rocesses.
\newblock {\em arXiv preprint arXiv:1708.06302}.

\bibitem[Kaufman et~al., 2008]{kaufman2008}
Kaufman, C.~G., Schervish, M.~J., and Nychka, D.~W. (2008).
\newblock {C}ovariance {T}apering for {L}ikelihood-{B}ased {E}stimation in
  {L}arge {S}patial {D}ata {S}ets.
\newblock {\em Journal of the American Statistical Association},
  103(484):1545--1555.

\bibitem[Kim et~al., 2004]{kim2004}
Kim, S.-Y., Lee, J.-T., Hong, Y.-C., Ahn, K.-J., and Kim, H. (2004).
\newblock {D}etermining the {T}hreshold {E}ffect of {O}zone on {D}aily
  {M}ortality: {A}n {A}nalysis of {O}zone and {M}ortality in {S}eoul, {K}orea,
  1995--1999.
\newblock {\em Environmental Research}, 94(2):113--119.

\bibitem[Kr{\"u}ger et~al., 2016]{kruger2016}
Kr{\"u}ger, F., Lerch, S., Thorarinsdottir, T.~L., and Gneiting, T. (2016).
\newblock {P}robabilistic {F}orecasting and {C}omparative {M}odel {A}ssessment
  {B}ased on {M}arkov {C}hain {M}onte {C}arlo {O}utput.
\newblock {\em arXiv preprint arXiv:1608.06802}.

\bibitem[Lippmann, 1989]{lippmann1989}
Lippmann, M. (1989).
\newblock {H}ealth {E}ffects of {O}zone a {C}ritical {R}eview.
\newblock {\em Japca}, 39(5):672--695.

\bibitem[Loomis et~al., 1999]{loomis1999}
Loomis, D., Castillejos, M., Gold, D.~R., McDonnell, W., and Borja-Aburto,
  V.~H. (1999).
\newblock {A}ir {P}ollution and {I}nfant {M}ortality in {M}exico {C}ity.
\newblock {\em Epidemiology}, pages 118--123.

\bibitem[MacKay, 1998]{mackay1998}
MacKay, D.~J. (1998).
\newblock {I}ntroduction to {G}aussian {P}rocesses.
\newblock {\em NATO ASI Series F Computer and Systems Sciences}, 168:133--166.

\bibitem[Mage et~al., 1996]{mage1996}
Mage, D., Ozolins, G., Peterson, P., Webster, A., Orthofer, R., Vandeweerd, V.,
  and Gwynne, M. (1996).
\newblock {U}rban {A}ir {P}ollution in {M}egacities of the {W}orld.
\newblock {\em Atmospheric Environment}, 30(5):681--686.

\bibitem[Porcu et~al., 2016]{porcu2016}
Porcu, E., Bevilacqua, M., and Genton, M.~G. (2016).
\newblock {S}patio-{T}emporal {C}ovariance and {C}ross-{C}ovariance {F}unctions
  of the {G}reat {C}ircle {D}istance on a {S}phere.
\newblock {\em Journal of the American Statistical Association},
  111(514):888--898.

\bibitem[Prado and West, 2010]{prado2010}
Prado, R. and West, M. (2010).
\newblock {\em {T}ime {S}eries: {M}odeling, {C}omputation, and {I}nference}.
\newblock CRC Press.

\bibitem[Rasmussen and Williams, 2006]{rasmussen2006}
Rasmussen, C.~E. and Williams, C. K.~I. (2006).
\newblock {G}aussian {P}rocesses for {M}achine {L}earning.
\newblock {\em The MIT Press, Cambridge, MA}.

\bibitem[Riojas-Rodr{\'\i}guez et~al., 2014]{riojas2014}
Riojas-Rodr{\'\i}guez, H., {\'A}lamo-Hern{\'a}ndez, U., Texcalac-Sangrador,
  J.~L., and Romieu, I. (2014).
\newblock {H}ealth {I}mpact {A}ssessment of {D}ecreases in {PM10} and {O}zone
  {C}oncentrations in the {M}exico {C}ity {M}etropolitan {A}rea: {A} {B}asis
  for a {N}ew {A}ir {Q}uality {M}anagement {P}rogram.
\newblock {\em {S}alud {P}{\'u}blica de {M}{\'e}xico}, 56:579--591.

\bibitem[Romieu et~al., 1996]{romieu1996}
Romieu, I., Meneses, F., Ruiz, S., Sienra, J.~J., Huerta, J., White, M.~C., and
  Etzel, R.~A. (1996).
\newblock {E}ffects of {A}ir {P}ollution on the {R}espiratory {H}ealth of
  {A}sthmatic {C}hildren l{L}ving in {M}exico {C}ity.
\newblock {\em American {J}ournal of {R}espiratory and {C}ritical {C}are
  {M}edicine}, 154(2):300--307.

\bibitem[Sahu et~al., 2007]{sahu2007}
Sahu, S.~K., Gelfand, A.~E., and Holland, D.~M. (2007).
\newblock {H}igh-{R}esolution {S}pace--{T}ime {O}zone {M}odeling for
  {A}ssessing {T}rends.
\newblock {\em Journal of the American Statistical Association},
  102(480):1221--1234.

\bibitem[Salam et~al., 2005]{salam2005}
Salam, M.~T., Millstein, J., Li, Y.-F., Lurmann, F.~W., Margolis, H.~G., and
  Gilliland, F.~D. (2005).
\newblock {B}irth {O}utcomes and {P}renatal {E}xposure to {O}zone, {C}arbon
  {M}onoxide, and {P}articulate {M}atter: {R}esults from the {C}hildren’s
  {H}ealth {S}tudy.
\newblock {\em Environmental Health Perspectives}, 113(11):1638.

\bibitem[Schoenberg, 1942]{schoenberg1942}
Schoenberg, I.~J. (1942).
\newblock Positive {D}efinite {F}unctions on {S}pheres.
\newblock {\em Duke Math. Journal}, 9:96--108.

\bibitem[Shirota et~al., 2017]{shirota2017}
Shirota, S., Gelfand, A.~E., et~al. (2017).
\newblock {S}pace and {C}ircular {T}ime {L}og {G}aussian {C}ox {P}rocesses with
  {A}pplication to {C}rime {E}vent {D}ata.
\newblock {\em The Annals of Applied Statistics}, 11(2):481--503.

\bibitem[Shumway and Stoffer, 2017]{shumway2017}
Shumway, R.~H. and Stoffer, D.~S. (2017).
\newblock {\em Time Series Analysis and Its Applications}.
\newblock Springer.

\bibitem[Sillman, 1999]{sillman1999}
Sillman, S. (1999).
\newblock The {R}elation {B}etween {O}zone, {NOx} and {H}ydrocarbons in {U}rban
  and {P}olluted {R}ural {E}nvironments.
\newblock {\em Atmospheric Environment}, 33(12):1821--1845.

\bibitem[Solin and S{\"a}rkk{\"a}, 2014]{solin2014}
Solin, A. and S{\"a}rkk{\"a}, S. (2014).
\newblock {E}xplicit {L}ink between {P}eriodic {C}ovariance {F}unctions and
  {S}tate {S}pace {M}odels.
\newblock In {\em Artificial Intelligence and Statistics}, pages 904--912.

\bibitem[Stein, 1999]{stein1999}
Stein, M.~L. (1999).
\newblock {\em {I}nterpolation of {S}patial {D}ata: {S}ome {T}heory for
  {K}riging}.
\newblock Springer Science \& Business Media.

\bibitem[Stein, 2005]{stein2005}
Stein, M.~L. (2005).
\newblock {S}tatistical {M}ethods for {R}egular {M}onitoring {D}ata.
\newblock {\em Journal of the Royal Statistical Society: Series B (Statistical
  Methodology)}, 67(5):667--687.

\bibitem[Stein, 2008]{stein2008}
Stein, M.~L. (2008).
\newblock A {M}odeling {A}pproach for {L}arge {S}patial {D}atasets.
\newblock {\em Journal of the Korean Statistical Society}, 37(1):3--10.

\bibitem[Stein, 2014]{stein2014}
Stein, M.~L. (2014).
\newblock {L}imitations on {L}ow {R}ank {A}pproximations for {C}ovariance
  {M}atrices of {S}patial {D}ata.
\newblock {\em Spatial Statistics}, 8:1--19.

\bibitem[Stein et~al., 2004]{stein2004}
Stein, M.~L., Chi, Z., and Welty, L.~J. (2004).
\newblock {A}pproximating {L}ikelihoods for {L}arge {S}patial {D}ata {S}ets.
\newblock {\em Journal of the Royal Statistical Society: Series B (Statistical
  Methodology)}, 66(2):275--296.

\bibitem[Vecchia, 1988]{vecchia1988}
Vecchia, A.~V. (1988).
\newblock {E}stimation and {M}odel {I}dentification for {C}ontinuous {S}patial
  {P}rocesses.
\newblock {\em Journal of the Royal Statistical Society. Series B
  (Methodological)}, pages 297--312.

\bibitem[Weschler, 2006]{weschler2006}
Weschler, C.~J. (2006).
\newblock {O}zone’s {I}mpact on {P}ublic {H}ealth: {C}ontributions from
  {I}ndoor {E}xposures to {O}zone and {P}roducts of {O}zone-{I}nitiated
  {C}hemistry.
\newblock {\em Environmental Health Perspectives}, 114(10):1489.

\bibitem[West and Harrison, 1997]{west1997}
West, M. and Harrison, J. (1997).
\newblock {\em {B}ayesian {F}orecasting and {D}ynamic {M}odels}.
\newblock Springer-Verlag New York, Inc., New York, NY, USA.

\bibitem[White et~al., 2018]{white2018a}
White, P.~A., Gelfand, A.~E., Rodrigues, E.~R., and Tzintzun, G. (2018).
\newblock {P}ollution {S}tate {M}odeling for {M}exico {C}ity.
\newblock {\em arXiv preprint arXiv:1807.03935}.

\bibitem[White and Porcu, 2018]{white2018b}
White, P.~A. and Porcu, E. (2018).
\newblock {T}owards a {C}omplete {P}icture of {C}ovariance {F}unctions on
  {S}pheres {C}ross {T}ime.
\newblock {\em arXiv preprint arXiv:1807.04272}.

\end{thebibliography}

\newpage

\begin{table}[H]
\centering
\footnotesize
\begin{tabular}{|l|l|l|rrrrr|}
  \hline
Model & Equation & Space & ES & CRPS & MAPE & RMSPE & 90\% CVG \\
   \hline
1 & $\mathcal{M}_{\alpha_1,1/2}$ $\times$ $\mathcal{M}_{\alpha_2,1/2}$ $\times$ $\mathcal{M}_{\alpha_3,1/2}$& $ \mathcal{D} \times \S^1 \times \R $ & 21.518 & 3.530 & 4.752 & 6.611 & 0.914 \\ 
2 & (\ref{eq:space2}) & $( \mathcal{D}\times \R)$ & 24.889 & 4.139 & 5.645 & 7.554 & 0.932 \\
3 & $\mathcal{M}_{\alpha,1/2}$ $\times$  (\ref{eq:space2}) & $ ( \mathcal{D} \times \R ) \times \S^1$ & 19.941 & 3.238 & 4.531 & 6.085 & 0.903 \\  
4 &(\ref{eq:space4}) & $( \mathcal{D} \times \S^1)$ & 24.726 & 4.295 & 5.851 & 7.890 & 0.914 \\
5 & $\mathcal{M}_{\alpha,1/2}$ $\times$ (\ref{eq:space4}) & $(  \mathcal{D} \times \S^1) \times \R$ & 19.773 & 3.249 & 4.309 & 6.050 & 0.908 \\  
6 &(\ref{eq:time7}) $\times$ $\mathcal{M}_{\alpha,1/2}$ & ${\cal D} \times (\S^1 \times \R) $ & 21.045 & 3.452 & 4.635 & 6.514 & 0.911 \\ 
7 &(\ref{eq:time10}) $\times$ $\mathcal{M}_{\alpha,1/2}$ & $ {\cal D} \times (\S^1 \times \R)  $ & \textbf{19.652}  & \textbf{3.214} & 4.273 & 6.025 & 0.919 \\ 
   \hline
\end{tabular}
\caption{Model Comparison results for the Mexico City ozone data. Parentheses in the ``Space'' column indicate that those quantities are nonseparable in the covariance model. The lowest ES and CRPS are highlighted.  }\label{tab:space_time_comp}
\end{table}

\newpage

\begin{table}[H]
\centering
\begin{tabular}{lrrrr}
  \hline
 & Mean & Standard Deviation & 2.5\% & 97.5\% \\ 
  \hline
  $\beta_0$ & 7.3943 & 0.5356 & 6.0396 & 8.1924 \\ 
  $\beta_{RH}$ & -0.0207 & 0.0012 & -0.0230 & -0.0180 \\ 
  $\beta_{TMP}$ & 0.1129 & 0.0068 & 0.0994 & 0.1263 \\ 
  $c_t$ & 86.9006 & 4.9505 & 77.6969 & 97.4642 \\ 
  $c_s$ & 22.3190 & 0.7020 & 20.8030 & 23.6254 \\ 
  $\alpha$ & 0.6740 & 0.0095 & 0.6558 & 0.6940 \\ 
 $\tau^2$ & 0.0947 & 0.0019 & 0.0909 & 0.0985 \\ 
  $\sigma^2$ & 2.0981 & 0.0502 & 2.0015 & 2.1986 \\ 
  $\sigma^2 /(\sigma^2 + \tau^2)$ & 0.9568 & 0.0013 & 0.9542 & 0.9592 \\ 
   \hline
\end{tabular}
\caption{Posterior summaries for model parameters. }\label{tab:summaries}
\end{table}

\end{document}